\documentclass[runningheads]{llncs}
\usepackage{graphicx}

\usepackage{cite}
\usepackage[utf8]{inputenc}
\usepackage{quantikz}
\usepackage{hyperref}
\usepackage{amsmath}
\usepackage{amsfonts}
\usepackage{braket}
\usepackage{xcolor}
\usepackage{dsfont}
\usepackage{multirow}

\usepackage[]{changes}

\newsavebox{\boxA}
\newsavebox{\boxB}
\newsavebox{\boxC}

\usepackage{geometry}
\usepackage{color}

\hypersetup{hidelinks,colorlinks=true,breaklinks=true,urlcolor=blue,allcolors=blue,pdftitle={Title},pdfauthor={Author}}

\def\BibTeX{{\rm B\kern-.05em{\sc i\kern-.025em b}\kern-.08em
    T\kern-.1667em\lower.7ex\hbox{E}\kern-.125emX}}
    
\usepackage{hhline}
\usepackage{caption}
\usepackage{subcaption}
\usepackage{float}
\usepackage{tabularx}

\usepackage{changes}
\definechangesauthor[name="Ilya Safro", color=red]{is}
\definechangesauthor[name="Joey", color=orange]{jl}
\usepackage{hyperref}
\usepackage{todonotes}

\newif\ifarxiv
\arxivtrue

\newif\ificcs
\iccsfalse

\begin{document}
\title{Partitioning Dense Graphs with Hardware Accelerators}
%
%
\author{Xiaoyuan Liu\inst{1,2} \and
Hayato Ushijima-Mwesigwa\inst{2} \and
Indradeep Ghosh \inst{2} \and
Ilya Safro \inst{1}}

%
\authorrunning{X. Liu et al.}
%

\institute{University of Delaware,
Newark, DE, USA \email{\{joeyxliu,isafro\}@udel.edu} \and
Fujitsu Research of America, Inc., Sunnyvale, CA, USA
\email{\{hayato,ighosh\}@fujitsu.com}\\
}

\maketitle              
\begin{abstract}
Graph partitioning is a fundamental combinatorial optimization problem that attracts a lot of attention from theoreticians and practitioners due to its broad applications. From multilevel graph partitioning to more general-purpose optimization solvers such as Gurobi and CPLEX, a wide range of approaches have been developed. Limitations of these approaches are important to study in order to break the computational optimization barriers of this problem. As we approach the limits of Moore's law, there is now a need to explore ways of solving such problems with special-purpose hardware such as quantum computers or quantum-inspired accelerators. In this work, we experiment with solving the graph partitioning on the Fujitsu Digital Annealer (a special-purpose hardware designed for solving combinatorial optimization problems) and compare it with the existing top solvers. We demonstrate limitations of existing solvers on many dense graphs as well as those of the Digital Annealer on sparse graphs which opens an avenue to hybridize these approaches.

\end{abstract}

\section{Introduction}

There are several reasons to be optimistic about the future of quantum-inspired and quantum devices. However, despite their great potential, we also need to acknowledge that state-of-art classical methods are extremely powerful after years of relentless research and development. In classical computing, the development of algorithms, the rich mathematical framework behind them, and sophisticated data structures are relatively mature, whereas the area of quantum computing is still at its nascent stage. Many existing classical algorithms do not have provable or good enough bounds on the performance (e.g., they might not have ideal performance in the worst case), but in many applications, the worst-case scenarios are rather rarely seen. As a result, such algorithms, many of which heuristics, can achieve excellent results in terms of the solution quality or speed. Therefore, when utilizing emerging technologies such as quantum-inspired hardware accelerators and quantum computers to tackle certain problems, it is important to compare them not only with possibly slow but provably strong algorithms but also with the heuristic algorithms that exhibit reasonably good results on the instances of interest.

The graph partitioning \cite{bulucc2016recent} is one of the combinatorial optimization problems for which there exists a big gap between rigorous theoretical approaches that ensure best known worst-case scenarios, and heuristics that are designed to cope with application instances exhibiting a reasonable quality-speed trade-off. Instances that arise in practical applications often contain special structures on which heuristics are engineered and tuned. Because of its practical importance, a huge amount of work has been done for a big class of graphs that arise in such areas as combinatorial scientific computing, machine learning, bioinformatics, and social science, namely, \emph{sparse graphs}. Over the years, there were several benchmarks on which the graph partitioning algorithms have been tested and compared with each other to mention just a few  \cite{Walshaw09archive,davis2011university,bader201110th}. However, \emph{dense graphs} can be rarely found in them. 
\ifarxiv
The situation with general dense linear algebra instances, many of which are used to test graph partitioners, is just a little bit better. In many cases, working with dense graphs requires very different algorithms and advanced computational resources.
\fi
As a result, most existing excellent graph partitioning heuristics do not perform well in practice on dense graphs, while provable algorithms with complexity that depends on the number of edges (or non-zeros in the corresponding matrix) are extremely slow. As we also show in computational results, a  graph sparsification does not necessarily practically help to achieve high-quality solutions.

\paragraph{Multilevel Algorithms} This class of heuristics is one of the most successful for a variety of cut-based graph problems such as the minimum linear arrangement \cite{SafroRB06}, and vertex separator \cite{hager2018multilevel}. Specifically for a whole variety of (hyper)graph partitioning versions \cite{KarypisKumar99fast,karypis1998multilevel,amg-sss12,shaydulin2019algdist} these heuristics exhibit best quality/speed trade-off \cite{bulucc2016recent}. In multilevel graph partitioning frameworks, a hierarchy of coarse graph representations is constructed in such a way that each next coarser graph is smaller than the previous finer one, and a solution of the partitioning for the coarse graph can approximate that of the fine graph and be further improved using fast local refinement. Multilevel algorithms are ideally suited for sparse graphs and suffer from the same problems as the algebraic multigrid (which generalizes, to the best of our knowledge, all known multilevel coarsening for partitioning) on dense matrices. In addition, a real scalability of the existing refinement for partitioning is achieved only for sparse local problems. Typically, if the density is increasing throughout the hierarchy construction, various ad-hoc tricks are used to accelerate optimization sacrificing the solution quality. When such things happen at the coarse levels, an error is quickly accumulated. Here we compare our results with KaHIP \cite{SandersS13} which produced the best results among several multilevel solvers \cite{bulucc2016recent}.  

\paragraph{Hardware Accelerators for Combinatorial Problems} Hardware accelerators such as GPU have been pivotal in the recent advancements of fields such as machine learning. Due to the computing challenges arising as a result of the physical scaling limits of Moore's law, scientists have started to develop special-purpose hardware for solving combinatorial optimization problems. Examples of such hardware include adiabatic quantum computers \cite{johnson2011quantum}, complementary metal-oxide-semiconductor (CMOS) annealers \cite{aramon2019physics} and coherent Ising machines \cite{inagaki2016coherent}. The gate-based universal quantum computers can also be used to solve such optimization problems\cite{liu2021layer}. These novel technologies are all unified by an ability to solve the Ising model or, equivalently, the quadratic unconstrained binary optimization (QUBO) problem. The general QUBO is  NP-hard and many problems can be formulated as QUBO \cite{lucas2014ising}. 
\ifarxiv
Previous work on using QUBO based models include areas such as clustering and community detection \cite{negre2020detecting,cohen2021unified,kalehbasti2021ising,cohen2020ising}
\cite{shaydulin2018community,shaydulin2019network,ushijima2017graph},  chemistry \cite{hernandez2017enhancing,hernandez2016novel,terry2019quantum}, finance \cite{rosenberg2016solving}, and machine learning \cite{crawford2016reinforcement,henderson2018leveraging,khoshaman2018quantum,levit2017free}.
\fi
It is also often used as a subroutine to model large neighborhood local search \cite{liu2019leveraging}. The Fujitsu Digital Annealer (DA) \cite{digitalannealer},  used in this work, utilizes application-specific integrated circuit hardware for solving fully connected QUBO problems. Internally the hardware runs a modified version of the Metropolis-Hastings algorithm for simulated annealing. The hardware utilizes massive parallelization and a novel sampling technique. The novel sampling technique speeds up the traditional Markov Chain Monte Carlo by almost always moving to a new state instead of being stuck in a local minimum. Here, we use the third generation DA, which is a hybrid software-hardware configuration that supports up to 100,000 binary variables. DA also supports users to specify inequality constraints and special equality constraints such as 1-hot and 2-way 1-hot constraints.

\paragraph{Our contribution}
The goal of this paper is twofold. First, we demonstrate that existing scalable graph partitioning dedicated solvers are struggling with the dense graphs not only in comparison to the special-purpose hardware accelerators but even sometimes if compared to generic global optimization solvers that are not converged. At the same time, we demonstrate a clear superiority of classical dedicated graph partitioning solvers on sparse instances. Second, this work is a step towards investigating what kind of problems we can solve using combinatorial hardware accelerators. Can we find problems that are hard for existing methods, but can be solved more efficiently with novel hardware and specialized algorithms? As an example, we explore the performance of Fujitsu Digital Annealer (DA) on graph partitioning and compare it with general-purpose solver Gurobi, and also graph partitioning solver KaHIP. 

We do not attempt to achieve an advantage for every single instance, especially since at the current stage, the devices we have right now are still facing many issues on scalability, noise, and so on. However, we advocate that hybridization of classical algorithms and specialized hardware (e.g., future quantum and existing quantum-inspired hardware) is a good candidate to break the barriers of the existing quality/speed trade-off.

\section{Graph Partitioning Formulations}
Let $G = (V, E)$ be an undirected, unweighted graph, where $V$ denotes the set of $n$ vertices, and $E$ denotes the set of $m$ edges. The goal of perfect balanced $k$-way graph partitioning (GP), is to partition $V$ into $k$ parts, $V_1, V_2, \cdots, V_k$, such that the $k$ parts are disjoint and have equal size, while minimizing the total number of \emph{cut edges}. A \emph{cut edge} is an edge that has two end vertices assigned to different parts. Sometimes, the quality of the partition can be improved if we allow some imbalance between different parts. In this case, we allow some imbalance factor $\epsilon > 0$, and each part can have at most $(1+\epsilon)\lceil n/k\rceil$ vertices.

\paragraph{Binary Quadratic Programming Formulation of GP}
We first review the integer quadratic programming formulation for $k$-way GP \cite{hager1999graph,ushijima2017graph}. When $k = 2$, we introduce binary variables $x_i \in \{0, 1\}$ for each vertex $i\in V$, where $x_i = 1$ if vertex $i$ is assigned to one part, and 0 otherwise. We denote by $\mathbf{x}$ the column vector $\mathbf{x} = (x_1, x_2, \cdots, x_n)^T$. The quadratic programming is then given by \begin{equation}\label{eq:quadprog}
\min_{\mathbf{x}}  \mathbf{x}^TL\mathbf{x} ~~~\text{such that }  x_i \in \{0, 1\}, ~\forall i\in V,
\end{equation}
where $L$ is the Laplacian matrix of graph $G$. 
\ifarxiv
The Laplacian matrix $L$ is defined as $L = D - A$, where $D$ is a diagonal matrix, with the degree of each node on the diagonal entries, and $A$ is the adjacency matrix of graph $G$, with $A_{ij} = 1, \forall (i, j)\in E$ and 0 otherwise. 
\fi
For perfect balance GP, we have the following equality constraint: \begin{eqnarray} \label{eq:constraint1}
\mathbf{x}^T\mathds{1} = \left\lceil\frac{n}{2}\right\rceil,
\end{eqnarray}
where $\mathds{1}$ is the column vector with ones. For the imbalanced case, we have the following inequality constraint: \begin{eqnarray}\label{eq:constraint2}
\mathbf{x}^T\mathds{1} \leq (1 + \epsilon)\left\lceil\frac{n}{2}\right\rceil.
\end{eqnarray}

When $k > 2$, we introduce binary variables $x_{i,j} \in \{0, 1\}$ for each vertex $i\in V$ and part $j$, where $x_{i,j} = 1$ if vertex $i$ is assigned to part $j$, and 0 otherwise. Let $\mathbf{x}_j$ denote the column vector $\mathbf{x}_j = (x_{1,j}, x_{2,j}, \cdots, x_{n,j})^T$ for $1 \leq j \leq k$. The quadratic programming formulation is then given by \begin{eqnarray*}
\min_{\mathbf{x}} && \frac{1}{2} \sum_{j=1}^k\mathbf{x}_j^TL\mathbf{x}_j \\
\text{s.t.} && \sum_{j=1}^k x_{i, j} = 1, ~~~\forall i \in V, \\
&& x_{i,j} \in \{0, 1\}, ~~~\forall i\in V, ~~~1 \leq j \leq k.
\end{eqnarray*}
Again, for perfect balance GP, we have another set of equality constraints: \begin{eqnarray*}
\mathbf{x}_j^T\mathds{1} = \left\lceil\frac{n}{k}\right\rceil, ~~~ 1 \leq j \leq k.
\end{eqnarray*}
For the imbalance case, we have the following inequality constraints: \begin{eqnarray*}
(1 - \epsilon)\left\lceil\frac{n}{k}\right\rceil \leq \mathbf{x}_j^T\mathds{1} \leq (1 + \epsilon)\left\lceil\frac{n}{k}\right\rceil, ~~~ 1 \leq j \leq k.
\end{eqnarray*}

\paragraph{QUBO Formulation}
To convert the problem into QUBO model, we will need to remove the constraints and add them as penalty terms to the objective function \cite{lucas2014ising}. For example, in the quadratic programming (\ref{eq:quadprog}) with the equality constraint (\ref{eq:constraint1}), we obtain the QUBO model as follows:\begin{eqnarray*}
\min_{\mathbf{x}} && \mathbf{x}^TL\mathbf{x} + P\left( \mathbf{x}^T\mathds{1} - \left\lceil \frac{n}{2}\right\rceil\right)^2 \\
\text{s.t.} && x_i \in \{0, 1\}, ~~~\forall i\in V,\nonumber
\end{eqnarray*}
where $P > 0$ is a postive parameter to penalize the violation of constraint (\ref{eq:constraint1}). For inequality constraints, we will introduce additional slack variables to first convert the inequality to equality constraints, and then add them as penalty terms to the objective function.

\section{Computational Experiments}
The goal of the experiments was to identify the class of instances that is more suitable to be solved using the QUBO framework and the current hardware.
We compare the performance of DA with exact solver Gurobi \cite{gurobi2018gurobi}, and the state-of-the-art multilevel graph partitioning solver KaHIP \cite{SandersS13}.
We set the time limit for DA and Gurobi to be 15 minutes. For KaHIP, we use KaFFPaE, a combination of distributed evolutionary algorithm and multilevel algorithm for GP. KaFFPaE computes partitions of very high quality when the imbalance factor $\epsilon > 0$, but does not perform very well for the perfectly balanced case when $\epsilon = 0$. Therefore we also enable KaBaPE, which is recommended by the developers. 
We run KaFFPaE with 24 processes in parallel, and set the time limit to be 30 minutes.

To evaluate the quality of the solution, we compare the approximation ratio, which is computed using the GP cut found by each solver divided by the best-known value. For some graphs, we have the best-known provided from the benchmark \cite{Walshaw09archive}, otherwise we use the best results found by the three solvers as the best known. Since this is a minimization problem, the minimum possible value of the approximation ratio is 1, the smaller the better. For each graph and each solver used, we also provide the objective function value, i.e., the number of cut edges. 

\begin{figure}[htbp]
\begin{subfigure}{.33\textwidth}
  \centering
  \includegraphics[width=1\linewidth]{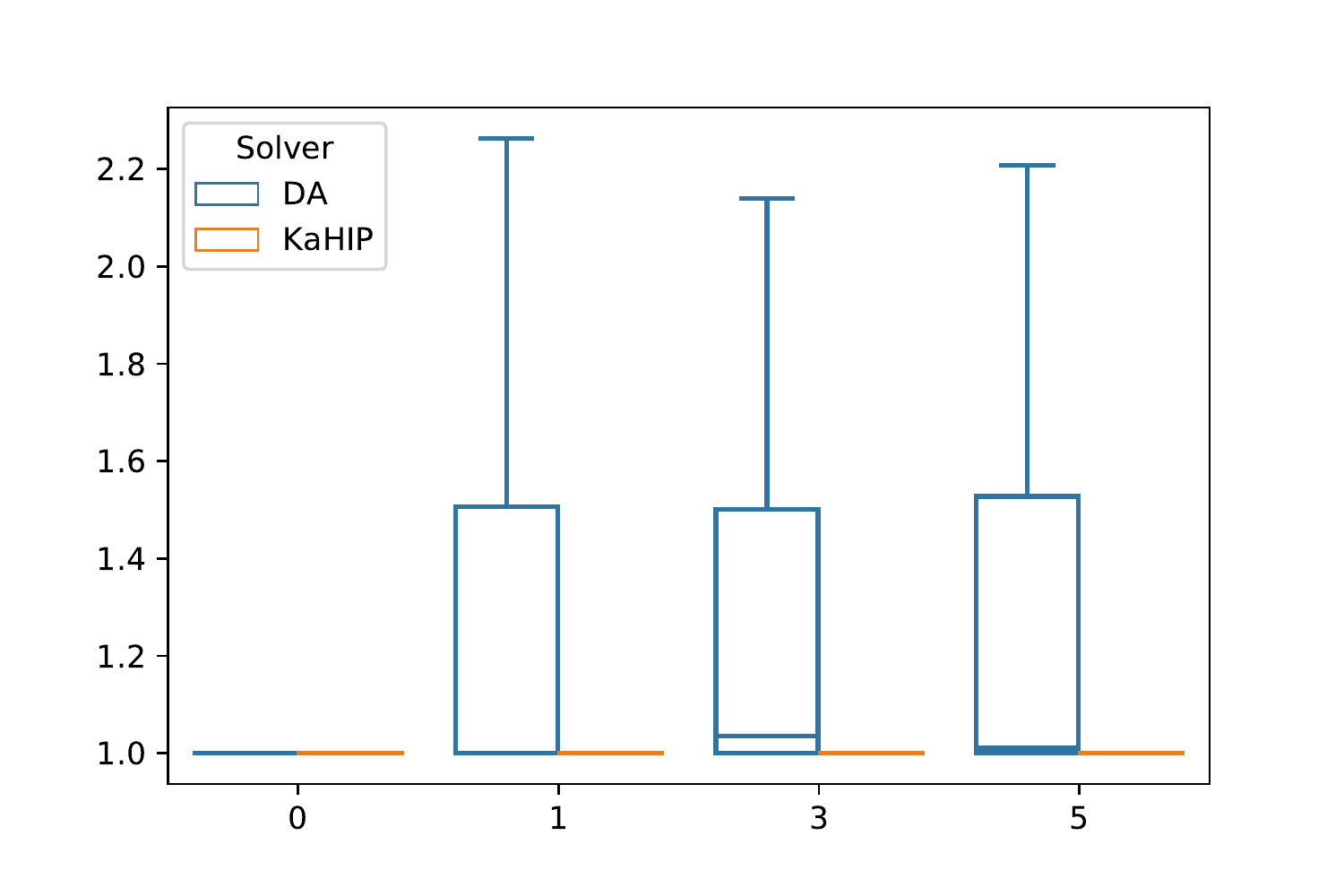}
  \subcaption{Walshaw $k=2$}
  \label{fig:sub-11}
\end{subfigure}
\begin{subfigure}{.33\textwidth}
  \centering
  \includegraphics[width=1\linewidth]{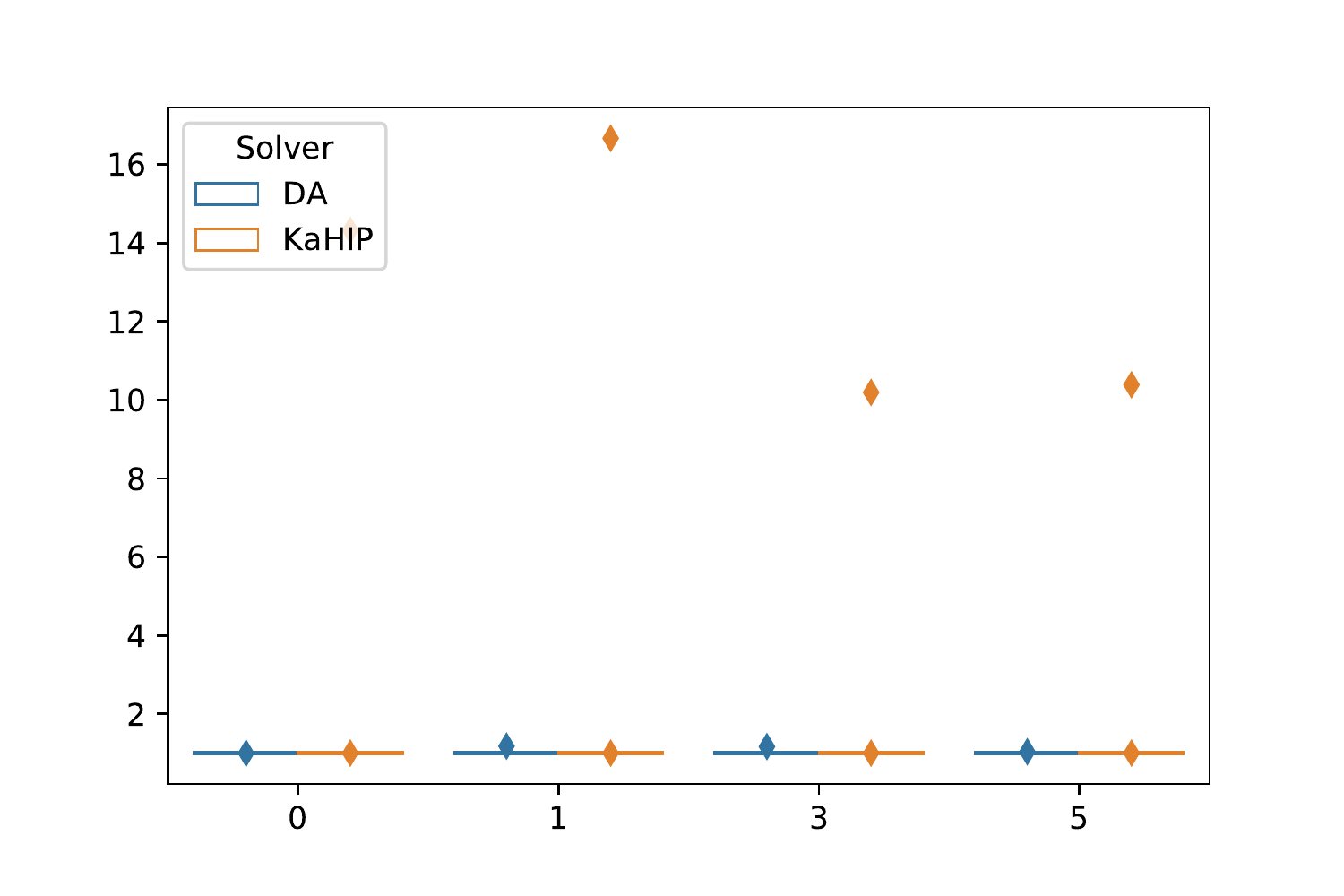}
  \caption{Suite Sparse $k=2$}
  \label{fig:sub-12}
\end{subfigure}
\begin{subfigure}{.33\textwidth}
  \centering
  \includegraphics[width=1\linewidth]{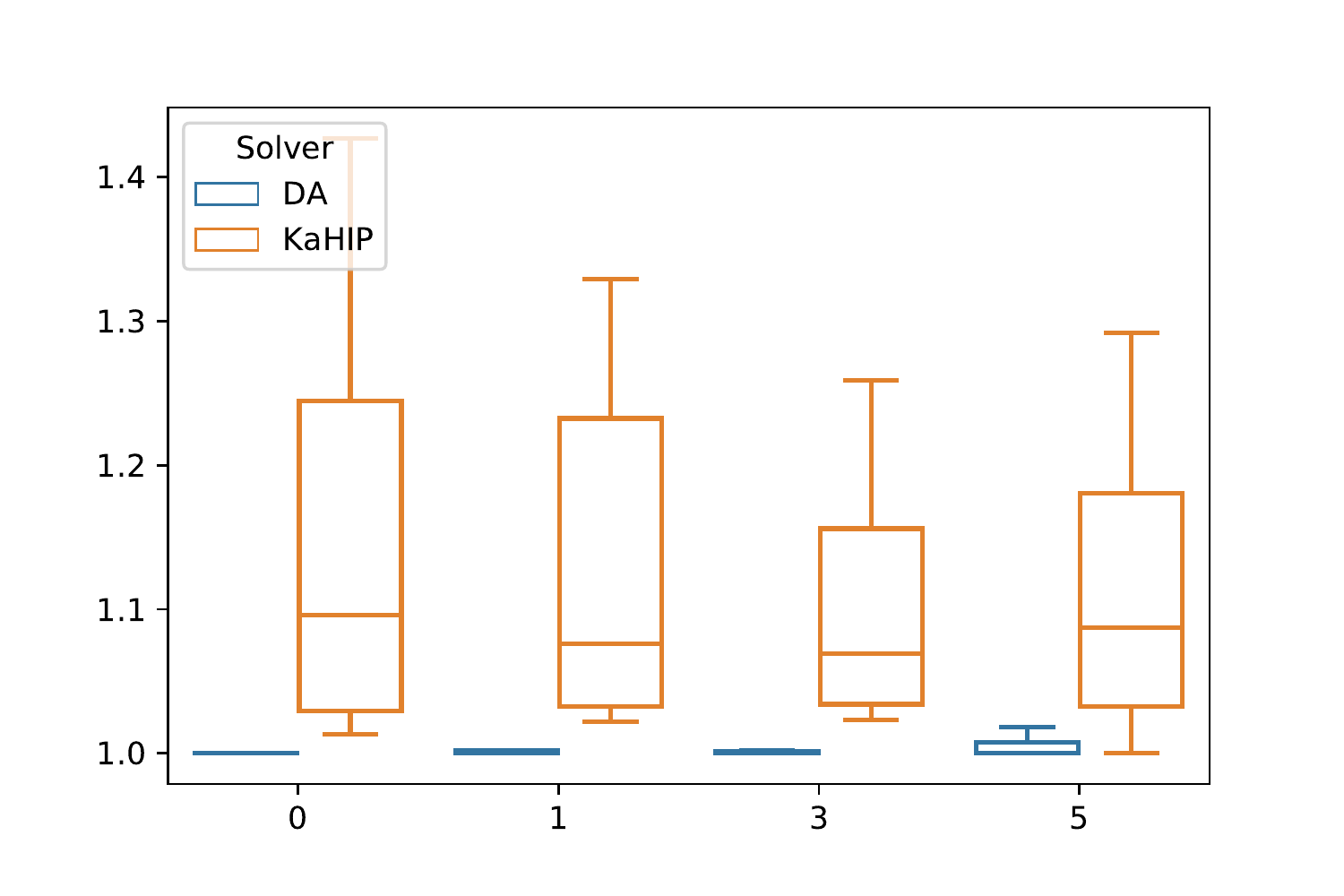}
  \caption{exdata $k=2$}
  \label{fig:sub-13}
\end{subfigure}
\begin{subfigure}{.33\textwidth}
  \centering
  \includegraphics[width=1\linewidth]{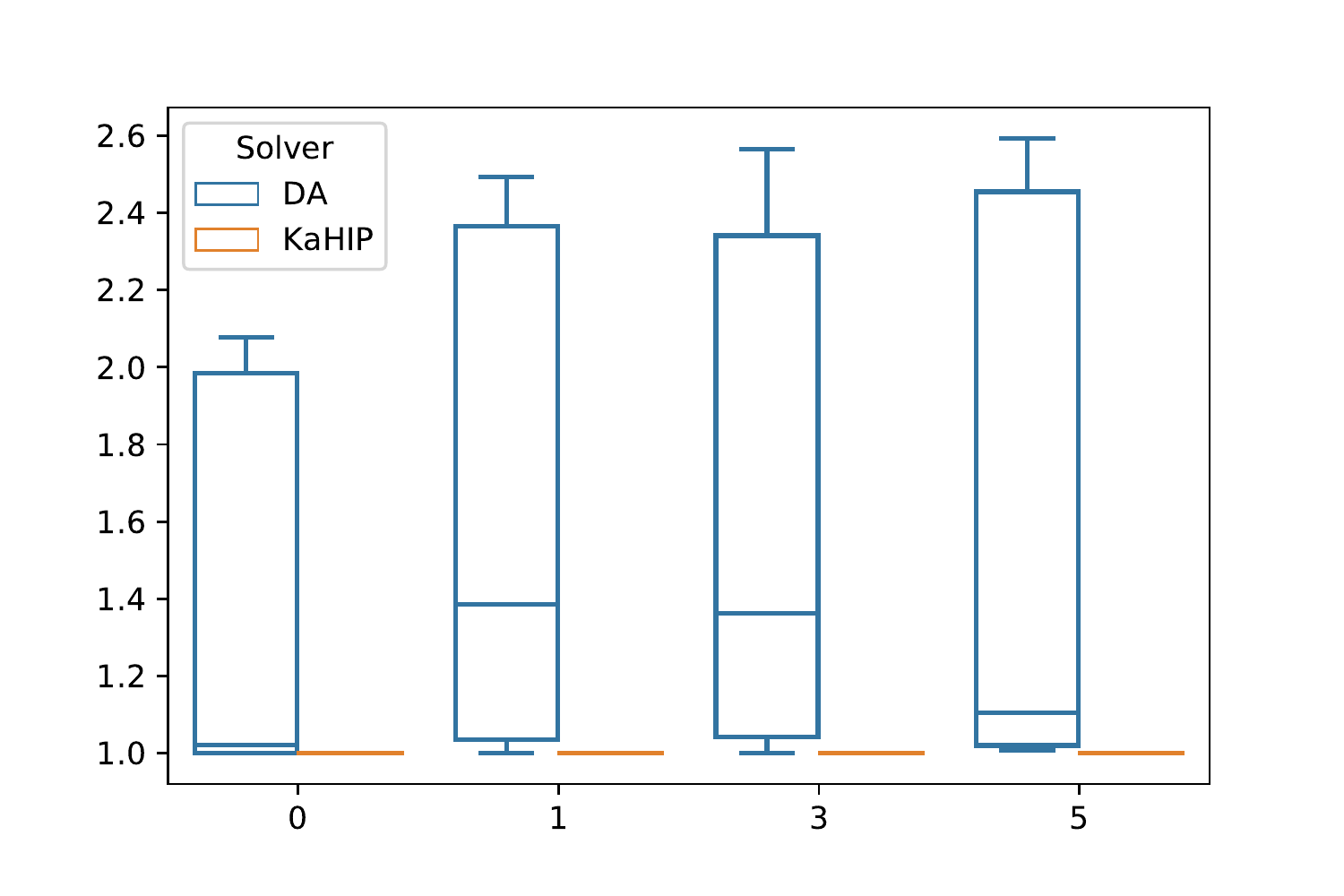}
  \caption{Walshaw $k=3$}
  \label{fig:sub-21}
\end{subfigure}
\begin{subfigure}{.33\textwidth}
  \centering
  \includegraphics[width=1\linewidth]{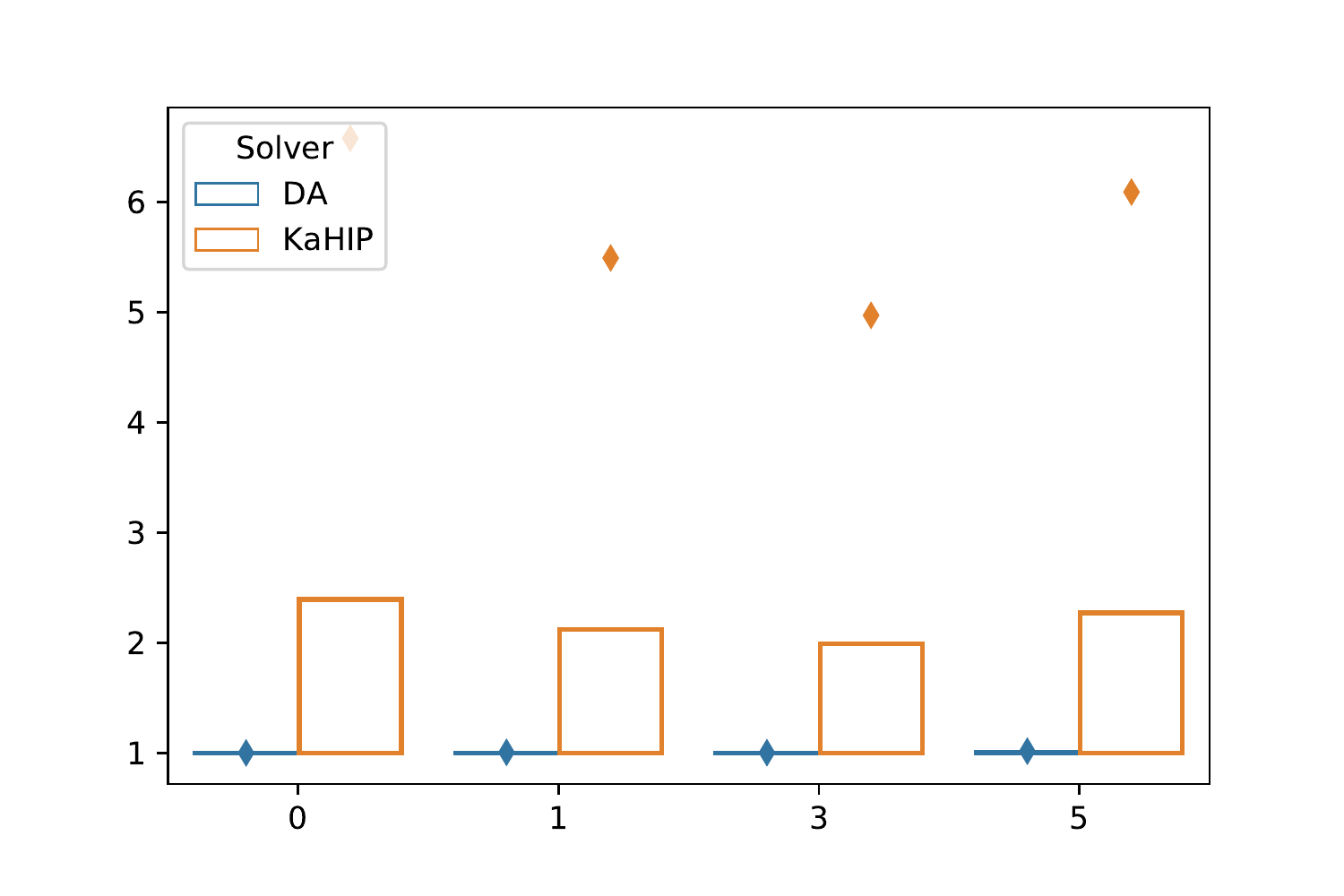}
  \caption{Suite Sparse $k=3$}
  \label{fig:sub-22}
\end{subfigure}
\begin{subfigure}{.33\textwidth}
  \centering
  \includegraphics[width=1\linewidth]{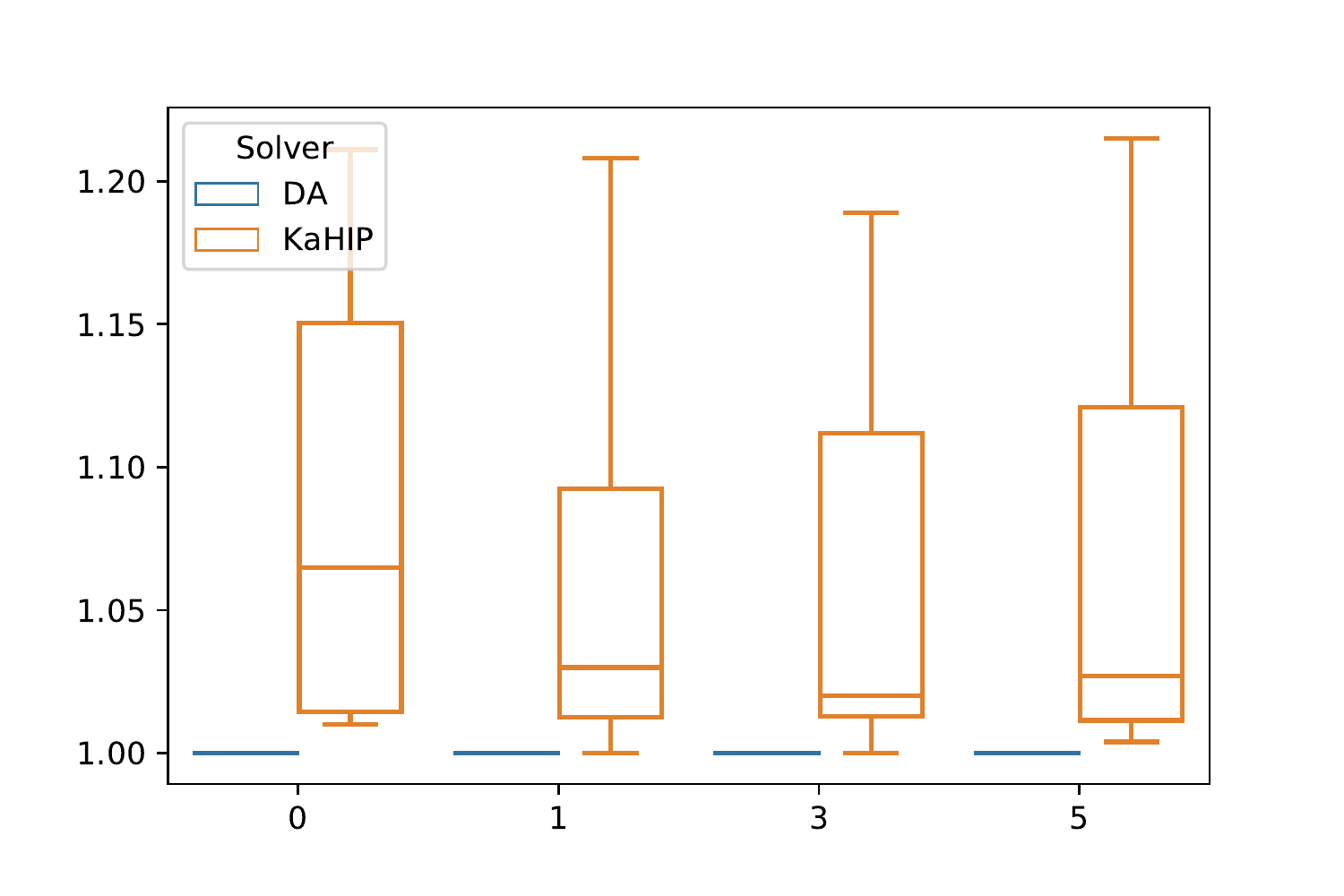}
  \caption{exdata $k=3$}
  \label{fig:sub-23}
\end{subfigure}
\begin{subfigure}{.33\textwidth}
  \centering
  \includegraphics[width=1\linewidth]{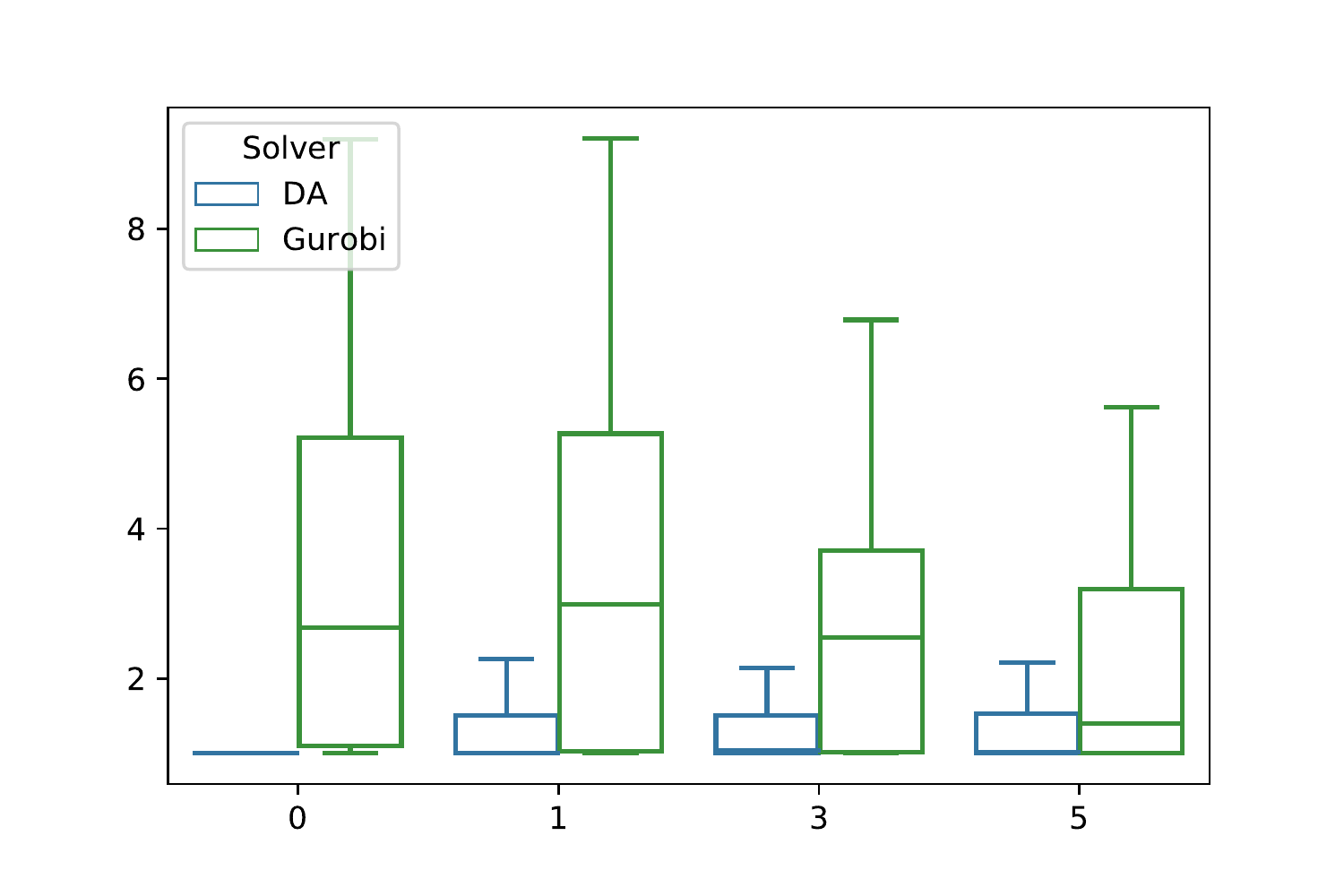}
  \caption{Walshaw $k=2$}
  \label{fig:sub-31}
\end{subfigure}
\begin{subfigure}{.33\textwidth}
  \centering
  \includegraphics[width=1\linewidth]{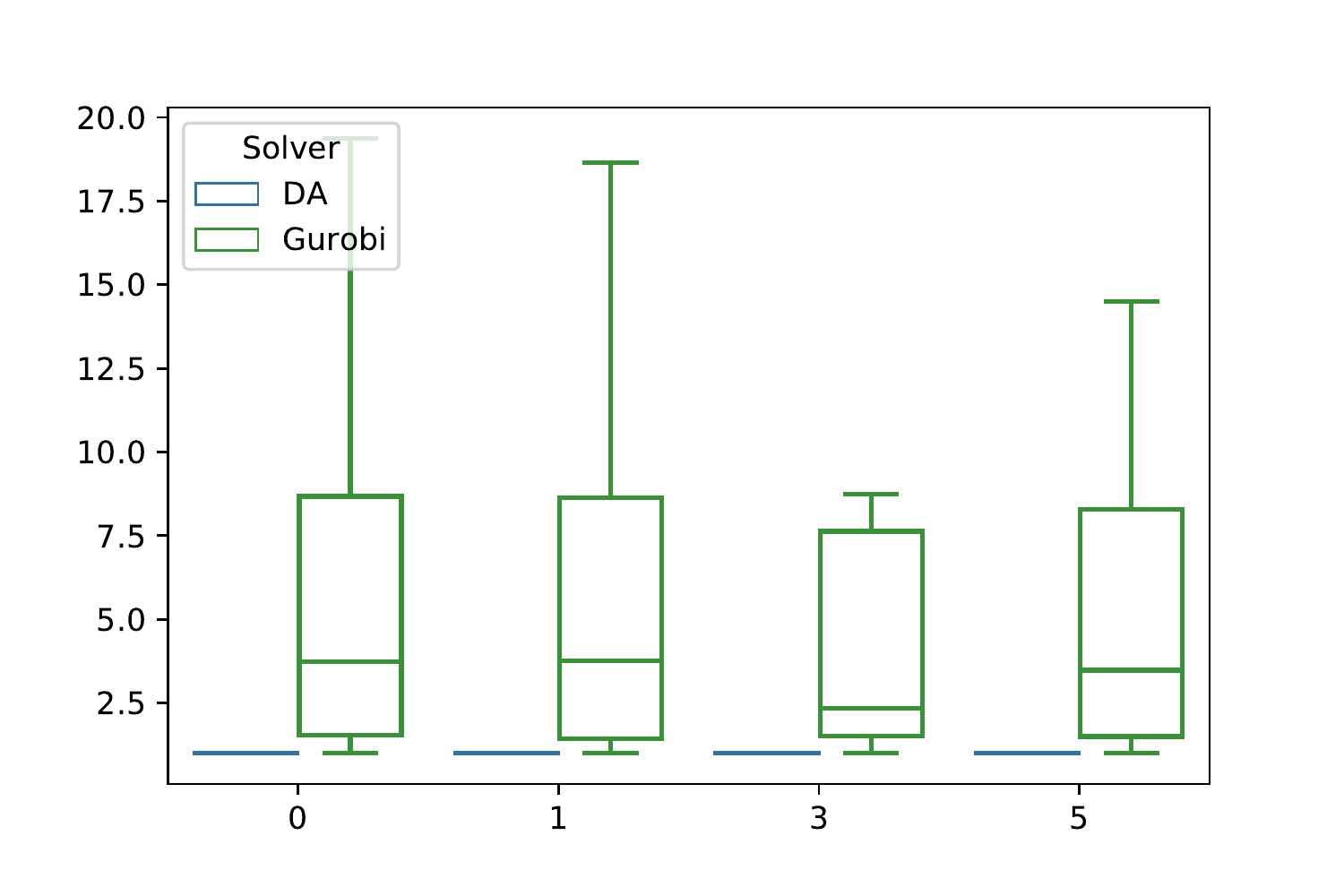}
  \caption{Suite Sparse $k=2$}
  \label{fig:sub-32}
\end{subfigure}
\begin{subfigure}{.33\textwidth}
  \centering
  \includegraphics[width=1\linewidth]{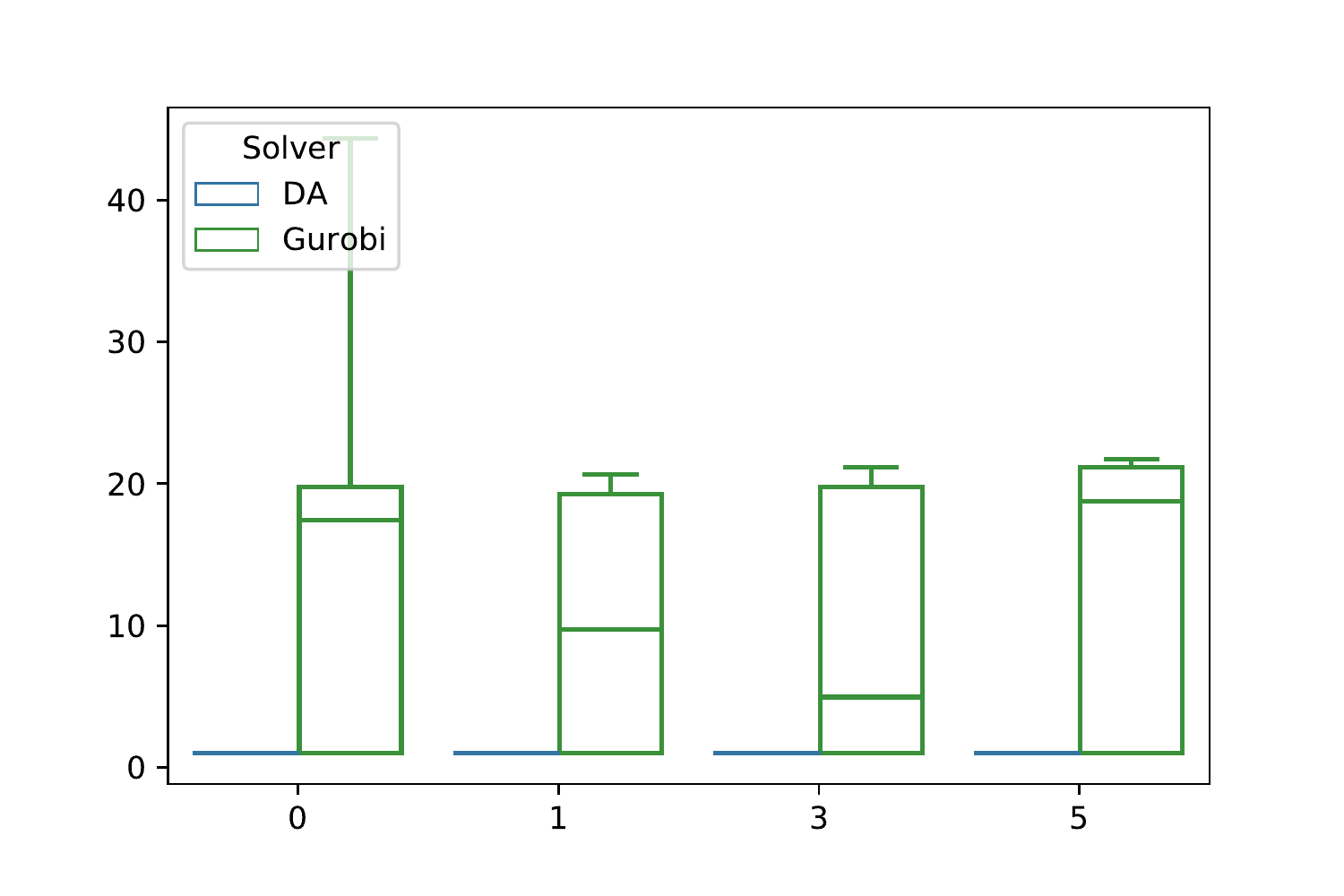}
  \caption{exdata $k=2$}
  \label{fig:sub-33}
\end{subfigure}
\begin{subfigure}{.33\textwidth}
  \centering
  \includegraphics[width=1\linewidth]{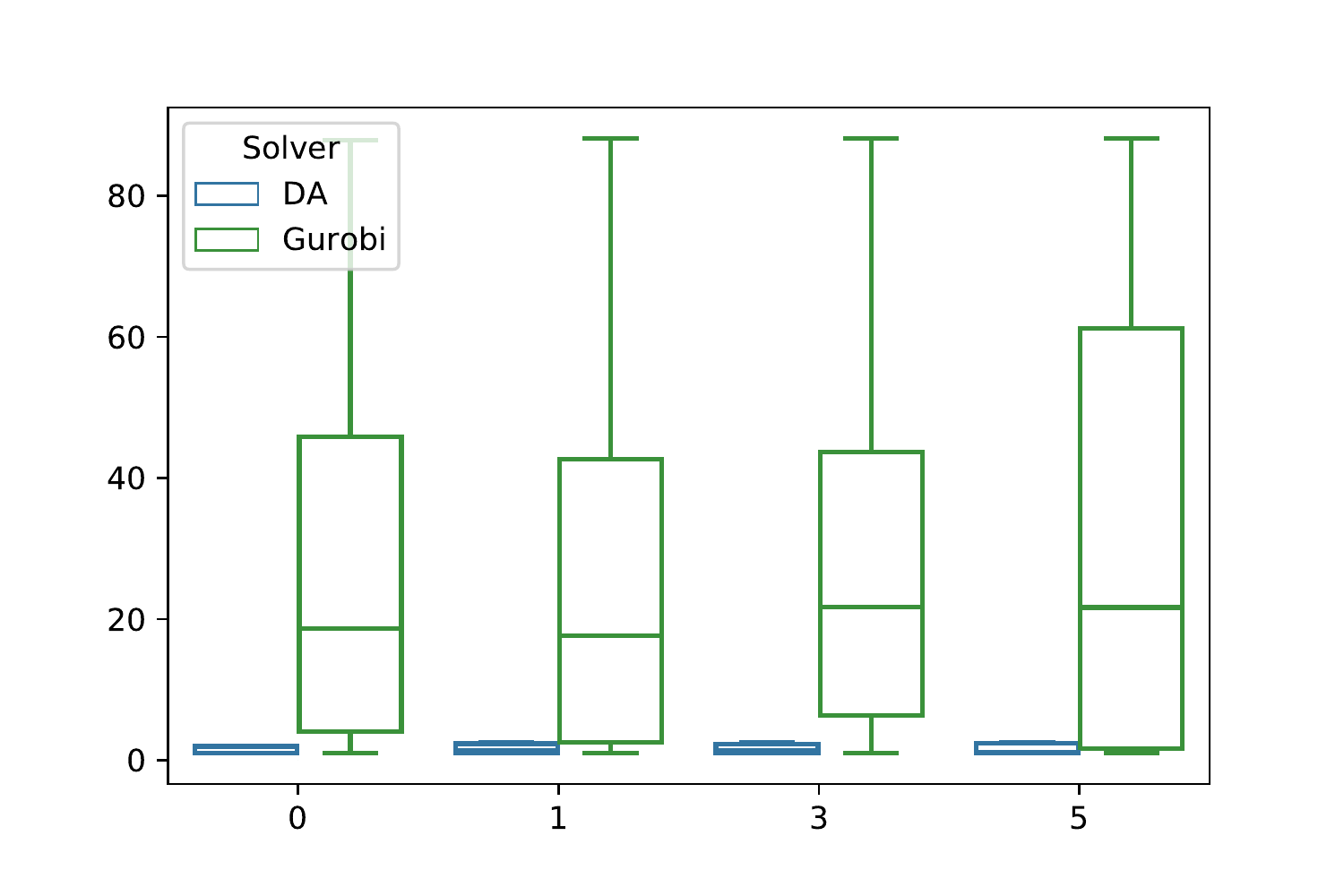}
  \caption{Walshaw $k=3$}
  \label{fig:sub-41}
\end{subfigure}
\begin{subfigure}{.33\textwidth}
  \centering
  \includegraphics[width=1\linewidth]{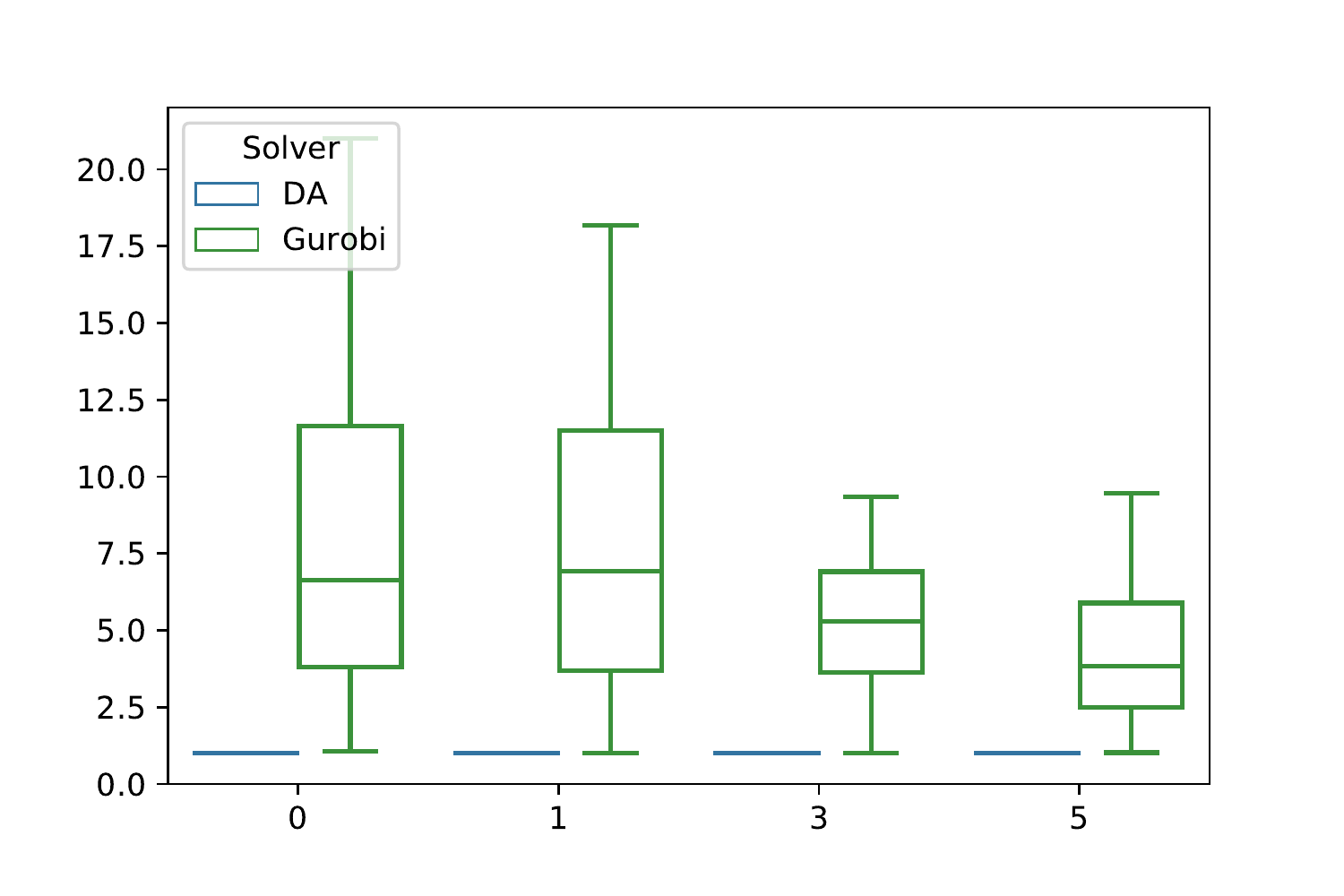}
  \caption{Suite Sparse $k=3$}
  \label{fig:sub-42}
\end{subfigure}
\begin{subfigure}{.33\textwidth}
  \centering
  \includegraphics[width=1\linewidth]{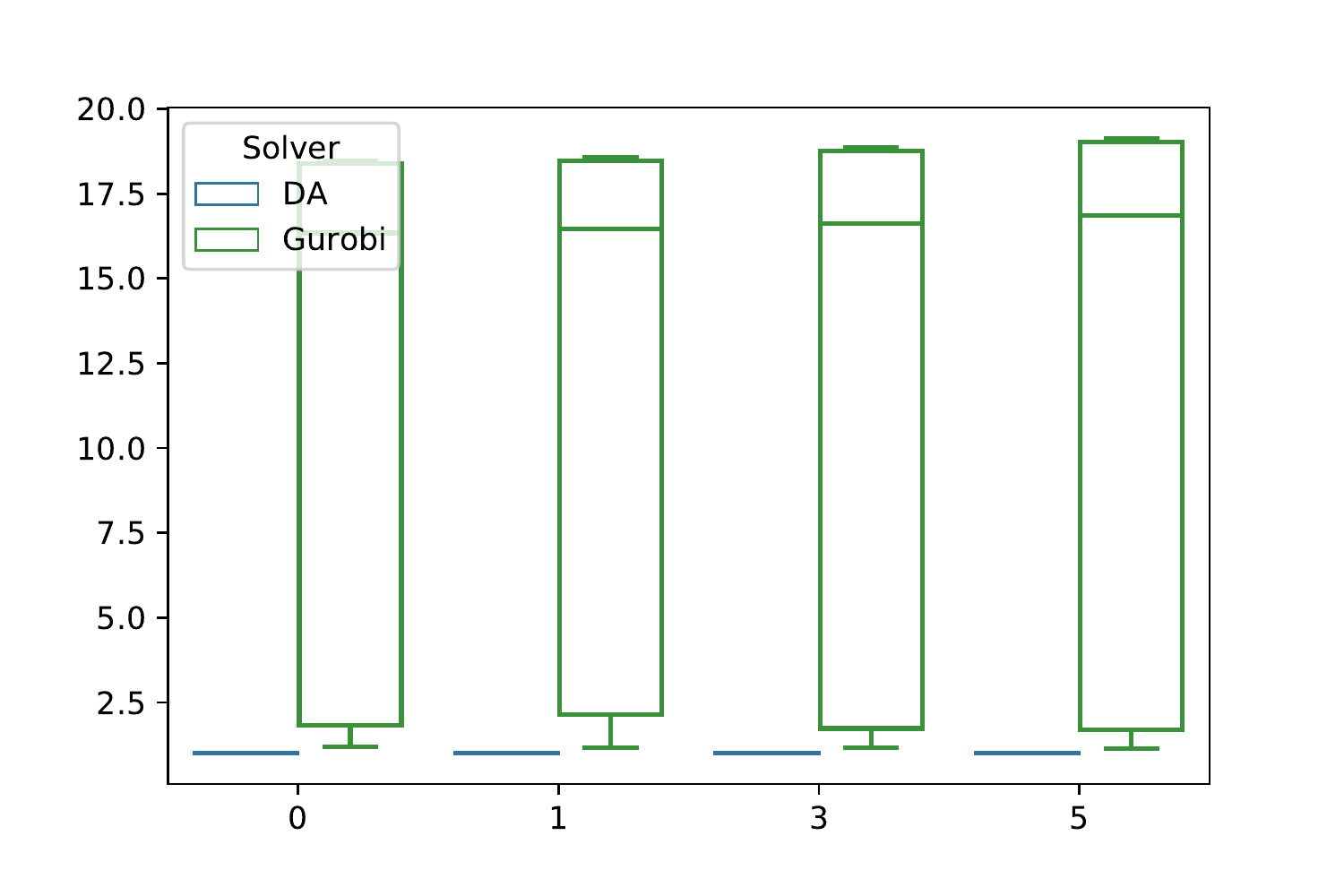}
  \caption{exdata $k=3$}
  \label{fig:sub-43}
\end{subfigure}
\caption{Comparison of DA with KaHIP (dedicated GP solver), and Gurobi (general-purpose solver) for sparse and dense graphs respectively. The y-axis represents the approximation ratio (solution to best-solution ratio), the minimum possible value of the approximation ratio is 1, the smaller the better. The x-axis represents the imbalance factor as percentage}
\label{fig:fig}
\end{figure}

\paragraph{Graph Partitioning on Sparse Graphs}
We first test the three solvers on instances from the Walshaw graph partitioning archive \cite{Walshaw09archive}. The information of the graphs is given in Table \ref{tab:gp_k_2_walshaw}, where $|V|$ is the number of nodes of the graph, and $d_{avg} = |E|/|V|$ describes the density of the graph. We present the summary of the results with box plots in Fig. \ref{fig:fig} (a), (d), (g) and (j). We also provide the objective function value of each graph obtained by the three solvers in Table \ref{tab:gp_k_2_walshaw}. We observe that in Figure \ref{fig:fig} (g) and (j), where we compare DA and Gurobi, DA can find the best-known partition for most instances, and perform better compared to Gurobi. However, for several sparse graphs, i.e., $d_{avg} < 3$, for example, \texttt{uk}, \texttt{add32} and \texttt{4elt}, DA can not find the best-known solutions. For these sparse graphs, multilevel graph partitioning solvers such as KaHIP can usually perform an effective coarsening and uncoarsening procedure based on local structures of the graph and therefore find good solutions quickly. As shown in Fig. \ref{fig:fig} (a), (d) and Table \ref{tab:gp_k_2_walshaw}, KaHIP performs better than DA. Based on the numerical results, we conclude that for the sparse graphs, generic and hardware QUBO solvers do not lead to many practical advantages. However, graphs with more complex structures, that bring practical challenges to the current solvers might benefit from using the QUBO and hardware accelerators.

\begin{table}[htbp]
\centering
\caption{Objective value (number of cut edges) of graphs from Walshaw graph partition archive\ifarxiv, $k=2, 3$.\fi}
\label{tab:gp_k_2_walshaw}
\begin{tabular}{|c||c|c||c|c|c|c|}
\hhline{-||--||----}
$k=2$ & \multirow{2}{*}{$|V|$} & \multirow{2}{*}{$d_{avg}$} & \multicolumn{4}{c|}{0\% imbalance} \\
\hhline{~||~~||----}
graph &&& Best known & DA & KaHIP & Gurobi\\
\hhline{-||--||----}
\texttt{add20} & 2395	&	3.12 & 596 & \textbf{596} & 613 & \textbf{596} \\
\hhline{-||--||----}
\texttt{data} & 2851	&	5.29	& 189 & \textbf{189} & \textbf{189} & 212 \\
\hhline{-||--||----}
\texttt{3elt} & 4720	&	2.91	& 90 & \textbf{90} & \textbf{90} & 91 \\
\hhline{-||--||----}
\texttt{uk} & 4824	&	1.42	&19 & \textbf{19} & \textbf{19} & \textbf{19} \\
\hhline{-||--||----}
\texttt{add32} & 4960	&	1.91	&11 & \textbf{11} & \textbf{11} & \textbf{11} \\
\hhline{-||--||----}
\texttt{bcsstk33} & 8738	&	33.37	& 10171 & \textbf{10171} & \textbf{10171} & 11674 \\
\hhline{-||--||----}
\texttt{whitaker3} & 9800	&	2.96	&127 & \textbf{127} & \textbf{127} & 640 \\
\hhline{-||--||----}
\texttt{crack} & 10240	&	2.97	& 184 & \textbf{184} & \textbf{184} & 331 \\
\hhline{-||--||----}
\texttt{wing\_nodal} & 10937	&	6.90	&1707 & \textbf{1707} & \textbf{1707} & 13739 \\
\hhline{-||--||----}
\texttt{fe\_4elt2} & 11143	&	2.95	&130 & 266 & \textbf{130} & 746 \\
\hhline{-||--||----}
\texttt{vibrobox} & 12328	&	13.40	&10343 & \textbf{10343} & \textbf{10343} & 31029 \\
\hhline{-||--||----}
\texttt{bcsstk29} & 13992	&	21.64	& 2843 & \textbf{2843} & \textbf{2843} & 26144 \\
\hhline{-||--||----}
\texttt{4elt} & 15606	&	2.94	& 139 & 256 & \textbf{139} & 592 \\
\hhline{-||--||----}
\texttt{fe\_sphere} & 16386	&	3.00	& 386 & \textbf{386} & \textbf{386} & 1082 \\
\hhline{-||--||----}
\texttt{cti} & 16840	&	2.86	& 334 & \textbf{334} & \textbf{334} & 2443 \\
\hhline{-||--||----}
\texttt{memplus} & 17758	&	3.05	& 5499 & 7286 & \textbf{5550} & 14030 \\
\hhline{-||--||----}
\end{tabular}
\begin{tabular}{|c|c|c|c|}
\hline
\multicolumn{4}{|c|}{1\% imbalance} \\ \hline
 Best known & DA & KaHIP & Gurobi\\
\hline
 585 & 586 & 591 & \textbf{585} \\
\hline
 188 & \textbf{188} & \textbf{188} & 193 \\
\hline
 89 & \textbf{89} & \textbf{89} & 90 \\
\hline
 19 & 43 & \textbf{19} & \textbf{19} \\
\hline
10 & 41 & \textbf{10} & \textbf{10} \\
\hline
 10097 & \textbf{10097} & \textbf{10097} & 53045 \\
\hline
 126 & \textbf{126} & \textbf{126} & 669 \\
\hline
 183 & 195 & \textbf{183} & 336 \\
\hline
 1695 & \textbf{1695} & \textbf{1695} & 10274 \\
\hline
 130 & 330 & \textbf{130} & 629 \\
\hline
 10310 & \textbf{10310} & \textbf{10310} & 31598 \\
\hline
 2818 & 2826 & \textbf{2818} & 25966 \\
\hline
 138 & 278 & \textbf{138} & 564 \\
\hline
 386 & \textbf{386} & \textbf{386} & 1128 \\
\hline
 318 & \textbf{318} & \textbf{318} & 2471 \\
\hline
 5452 & 7293 & \textbf{5476} & 11249 \\
\hline
\end{tabular}

\ifarxiv
\begin{tabular}{|c||c|c||c|c|c|c|}
\multicolumn{7}{c}{~}\\
\hhline{-||--||----}
$k=2$ & \multirow{2}{*}{$|V|$} & \multirow{2}{*}{$d_{avg}$} & \multicolumn{4}{c|}{3\% imbalance} \\
\hhline{~||~~||----}
graph &&& Best known & DA & KaHIP & Gurobi\\
\hhline{-||--||----}
\texttt{add20} & 2395	&	3.12	& 560 & \textbf{560} & 568 & \textbf{560} \\
\hhline{-||--||----}
\texttt{data} & 2851	&	5.29	& 185 & \textbf{185} & \textbf{185} & 212 \\
\hhline{-||--||----}
\texttt{3elt} & 4720	&	2.91	& 87 & \textbf{87} & \textbf{87} & \textbf{87} \\
\hhline{-||--||----}
\texttt{uk} & 4824	&	1.42	& 18 & 36 & \textbf{18} & \textbf{18} \\
\hhline{-||--||----}
\texttt{add32} & 4960	&	1.91	& 10 & 70 & \textbf{10} & \textbf{10} \\
\hhline{-||--||----}
\texttt{bcsstk33} & 8738	&	33.37	& 10064 & \textbf{10064} & \textbf{10064} & 43173 \\
\hhline{-||--||----}
\texttt{whitaker3} & 9800	&	2.96	& 126 & \textbf{126} & \textbf{126} & 195 \\
\hhline{-||--||----}
\texttt{crack} & 10240	&	2.97	& 182 & 195 & \textbf{182} & 571 \\
\hhline{-||--||----}
\texttt{wing\_nodal} & 10937	&	6.90	& 1678 & \textbf{1678} & \textbf{1678} & 11392 \\
\hhline{-||--||----}
\texttt{fe\_4elt2} & 11143	&	2.95	& 130 & 261 & \textbf{130} & 132 \\
\hhline{-||--||----}
\texttt{vibrobox} & 12328	&	13.40	& 10310 & \textbf{10310} & \textbf{10310} & 34490 \\
\hhline{-||--||----}
\texttt{bcsstk29} & 13992	&	21.64	& 2818 & 3273 & \textbf{2818} & 25601 \\
\hhline{-||--||----}
\texttt{4elt} & 15606	&	2.94	& 137 & 293 & \textbf{137} & 482 \\
\hhline{-||--||----}
\texttt{fe\_sphere} & 16386	&	3.00	& 384 & \textbf{384} & \textbf{384} & 1120 \\
\hhline{-||--||----}
\texttt{cti} & 16840	&	2.86	& 318 & 343 & \textbf{318} & 1424 \\
\hhline{-||--||----}
\texttt{memplus} & 17758	&	3.05	& 5352 & 7137 & \textbf{5362} & 11676 \\
\hhline{-||--||----}
\end{tabular}
\begin{tabular}{|c|c|c|c|}
 \multicolumn{4}{c}{~}\\
 \hhline{-||---}
\multicolumn{4}{|c|}{5\% imbalance} \\ \hline
 Best known & DA & KaHIP & Gurobi\\
\hline
536 & \textbf{536} & 540 & \textbf{536} \\
\hline
 181 & 188 & \textbf{181} & 190 \\
\hline
87 & \textbf{87} & \textbf{87} & \textbf{87} \\
\hline
 18 & 38 & \textbf{18} & \textbf{18} \\
\hline
 10 & 49 & 10 & 10 \\
\hline
 9914 & \textbf{9914} & 10010 & 11105 \\
\hline
 126 & \textbf{126} & \textbf{126} & 176 \\
\hline
 182 & 201 & \textbf{182} & 254 \\
\hline
 1668 & \textbf{1668} & \textbf{1668} & 9379 \\
\hline
  130 & 287 & 137 & \textbf{130} \\
\hline
 10310 & \textbf{10310} & \textbf{10310} & 31173 \\
\hline
 2818 & 2823 & \textbf{2818} & 25084 \\
\hline
 137 & 360 & \textbf{137} & 507 \\
\hline
 384 & \textbf{384} & \textbf{384} & 762 \\
\hline
 318 & 324 & \textbf{318} & 1455 \\
\hline
 5253 & 7001 & \textbf{5274} & 10923 \\
\hline
\end{tabular}

\begin{tabular}{|c||c|c|c|}
\multicolumn{4}{c}{~}\\
 \hhline{-||---}
$k=3$ & \multicolumn{3}{c|}{0\% imbalance} \\
\hhline{~||---}
graph &  DA & KaHIP & Gurobi\\
\hhline{-||---}
\texttt{add20} & \textbf{936} & 949 & 1148 \\
\hhline{-||---}
\texttt{data} & \textbf{261} & \textbf{261} & 5129 \\
\hhline{-||---}
\texttt{3elt} & 168 & \textbf{162} & 5023 \\
\hhline{-||---}
\texttt{uk} & 175 & \textbf{32} & 50 \\
\hhline{-||---}
\texttt{add32} & 133 & \textbf{28} & \textbf{28} \\
\hhline{-||---}
\texttt{bcsstk33} & \textbf{16247} & \textbf{16247} & 189894 \\
\hhline{-||---}
\texttt{whitaker3} & \textbf{253} & \textbf{253} & 12842 \\
\hhline{-||---}
\texttt{crack} & 492 & \textbf{288} & 20263 \\
\hhline{-||---}
\texttt{wing\_nodal} & 2864 & \textbf{2850} & 50296 \\
\hhline{-||---}
\texttt{fe\_4elt2} & 515 & \textbf{248} & 21800 \\
\hhline{-||---}
\end{tabular}
\begin{tabular}{|c|c|c|}
\multicolumn{3}{c}{~}\\
 \hline
 \multicolumn{3}{|c|}{1\% imbalance} \\
 \hline
  DA & KaHIP & Gurobi\\
\hline
 941 & \textbf{937} & 1093 \\
\hline
 \textbf{259} & \textbf{259} & 4539 \\
\hline
 185 & \textbf{162} & 4292 \\
\hline
 210 & \textbf{31} & 154 \\
\hline
 135 & \textbf{25} & \textbf{25} \\
\hline
 16155 & \textbf{16059} & 27722 \\
\hline
 412 & \textbf{253} & 12160 \\
\hline
 570 & \textbf{287} & 20304 \\
\hline
 3187 & \textbf{2844} & 50370 \\
\hline
 618 & \textbf{248} & 21865 \\
\hline
\end{tabular}
\begin{tabular}{|c|c|c|}
\multicolumn{3}{c}{~}\\
 \hline
 \multicolumn{3}{|c|}{3\% imbalance} \\
 \hline
DA & KaHIP & Gurobi\\
\hline
 924 & \textbf{923} & 1067 \\
\hline
 265 & \textbf{255} & 6528 \\
\hline
 177 & \textbf{162} & 4855 \\
\hline
 233 & \textbf{31} & 144 \\
\hline
 184 & \textbf{22} & \textbf{22} \\
\hline
 15960 & \textbf{15819} & 178712 \\
\hline
 422 & \textbf{253} & 12198 \\
\hline
 459 & \textbf{281} & 20304 \\
\hline
 2978 & \textbf{2828} & 50370 \\
\hline
 636 & \textbf{248} & 21865 \\
\hline
\end{tabular}
\begin{tabular}{|c|c|c|}
\multicolumn{3}{c}{~}\\
 \hline
 \multicolumn{3}{|c|}{5\% imbalance} \\
 \hline
  DA & KaHIP & Gurobi\\
\hline
 950 & \textbf{894} & 1029 \\
\hline
 253 & \textbf{250} & 6348 \\
\hline
 162 & \textbf{159} & 4449 \\
\hline
 202 & \textbf{29} & 33 \\
\hline
 172 & \textbf{19} & \textbf{19} \\
\hline
 15679 & \textbf{15552} & 47890 \\
\hline
 258 & \textbf{252} & 19294 \\
\hline
 573 & \textbf{281} & 20304 \\
\hline
 3228 & \textbf{2813} & 50370 \\
\hline
 643 & \textbf{248} & 21865 \\
\hline
\end{tabular}
\fi
\end{table}
\paragraph{Graph Partitioning on Dense Graphs}
To validate our conjecture, in the next set of experiments, we examine dense graphs from the  SuiteSparse Matrix Collection \cite{davis2011university} (see details in Table \ref{tab:gp_k_2_tamu}). The experimental results are presented in Fig. \ref{fig:fig} (b), (e), (h) and (k), and the objective function values of each graph obtained by the three solvers are provided in Table \ref{tab:gp_k_2_tamu}. We observe that for these dense graphs, in general, DA is able to find solutions that are usually at least as good as those produced by  KaHIP and Gurobi. In particular, we find that for one instance, \texttt{exdata\_1}, KaHIP fails significantly. We therefore use a graph generator MUSKETEER \cite{gutfraind2015multiscale} to generate similar instances\footnote{The \texttt{exdata} graph files are available here: \url{https://github.com/JoeyXLiu/dense-graph-exdata}}. \ifarxiv The details of the parameters used to generate the graphs can be found in the appendix. \fi In short, MUSKETEER applies perturbation to the original graph with a multilevel approach, the local editing preserves many network properties including different centralities measures, modularity, and clustering. The  information about generated instances is given in Table \ref{tab:gp_k_2_exdata}. The experiment results are presented in Fig. \ref{fig:fig} (c), (f), (i) and (l), and the objective function value of each graph obtained by the three solvers are provided in Table \ref{tab:gp_k_2_exdata}. We find that in most instances, DA outperforms KaHIP and Gurobi, demonstrating that in this class of problems, specialized hardware such as DA is having an advantage.
\begin{table*}[t]
\centering
\caption{Objective value (number of cut edges) of graphs from SuiteSparse Matrix Collection\ifarxiv, $k=2, 3$.\fi}
\label{tab:gp_k_2_tamu}
\begin{tabular}{|c||c|c||c|c|c||c|c|c|}
\hhline{-||--||---||---}
$k=2$ & \multirow{2}{*}{$|V|$} & \multirow{2}{*}{$d_{avg}$} & \multicolumn{3}{c||}{0\% imbalance} & \multicolumn{3}{c|}{1\% imbalance}\\
\hhline{~||~~||---||---}
graph &&& DA & KaHIP & Gurobi & DA & KaHIP & Gurobi\\
\hhline{-||--||---||---}
\texttt{exdata\_1} & 6001	&	188.59	& \textbf{2000} & 28646 & \textbf{2000} & \textbf{1980} & 33004 & \textbf{1980}\\
\hhline{-||--||---||---}
\texttt{TSC\_OPF\_1047} & 8140	&	123.39	& 1188 & \textbf{1187} & 1763 & 1381 & \textbf{1170} & 1586\\
\hhline{-||--||---||---}
\texttt{nd3k} & 9000	&	181.71	& \textbf{149880} & \textbf{149880} & 817727 & \textbf{149829} & \textbf{149829} & 818433\\
\hhline{-||--||---||---}
\texttt{nemeth26} & 9506	&	79.02	& \textbf{3298} & \textbf{3298} & 4842 & \textbf{3284} & \textbf{3284} & 3297\\
\hhline{-||--||---||---}
\texttt{mycielskian14} & 12287	&	150.38	& \textbf{553735} & \textbf{553735} & 924567 & \textbf{545355} & \textbf{545355} & 923991\\
\hhline{-||--||---||---}
\texttt{human\_gene2} & 14340	&	629.50	& \textbf{544938} & 546204 & 4514331 & \textbf{542850} & 543834 & 4514154\\
\hhline{-||--||---||---}
\texttt{opt1} & 15449	&	61.98	& \textbf{24725} & \textbf{24725} & 479054 & \textbf{24030} & \textbf{24030} & 448245\\
\hhline{-||--||---||---}
\texttt{gupta3} & 16783	&	277.26	& \textbf{1143782} & \textbf{1143782} & 2325983 & \textbf{1137072} & \textbf{1137072} & 2323035\\
\hhline{-||--||---||---}
\texttt{ramage02} & 16830	&	84.66	& \textbf{80940} & \textbf{80940} & 712906 & \textbf{80912} & \textbf{80912} & 707102\\
\hhline{-||--||---||---}
\texttt{pkustk07} & 16860	&	71.23	& \textbf{66852} & \textbf{66852} & 600140 & \textbf{66834} & \textbf{66834} & 588200\\
\hhline{-||--||---||---}
\end{tabular}

\ifarxiv
\begin{tabular}{|c||c|c||c|c|c||c|c|c|}
\multicolumn{9}{c}{~}\\
\hhline{-||--||---||---}
$k=2$ & \multirow{2}{*}{$|V|$} & \multirow{2}{*}{$d_{avg}$} & \multicolumn{3}{c||}{3\% imbalance} & \multicolumn{3}{c|}{5\% imbalance}\\
\hhline{~||~~||---||---}
graph &&& DA & KaHIP & Gurobi & DA & KaHIP & Gurobi\\
\hhline{-||--||---||---}
\texttt{exdata\_1} & 6001	&	188.59	& \textbf{1940} & 19776 & \textbf{1940} & \textbf{1900} & 19736 & \textbf{1900}\\
\hhline{-||--||---||---}
\texttt{TSC\_OPF\_1047} & 8140	&	123.39	& 1339 & \textbf{1146} & 1647 & 1185 & \textbf{1140} & 1591\\
\hhline{-||--||---||---}
\texttt{nd3k} & 9000	&	181.71	& \textbf{148935} & \textbf{148935} & 803125 & \textbf{148403} & \textbf{148403} & 727512\\
\hhline{-||--||---||---}
\texttt{nemeth26} & 9506	&	79.02	& \textbf{3284} & \textbf{3284} & 3290 & \textbf{3284} & \textbf{3284} & -1\\
\hhline{-||--||---||---}
\texttt{mycielskian14} & 12287	&	150.38	& \textbf{527882} & \textbf{527883} & 923991 & \textbf{509891} & \textbf{509891} & 923991\\
\hhline{-||--||---||---}
\texttt{human\_gene2} & 14340	&	629.50	& \textbf{538979} & 540182 & 4514154 & \textbf{533825} & 535779 & 4514154\\
\hhline{-||--||---||---}
\texttt{opt1} & 15449	&	61.98	& \textbf{22682} & \textbf{22682} & 59808 & \textbf{21685} & \textbf{21685} & 314167\\
\hhline{-||--||---||---}
\texttt{gupta3} & 16783	&	277.26	& \textbf{1122713} & \textbf{1122713} & 2293610 & \textbf{1107794} & \textbf{1107794} & 2284264\\
\hhline{-||--||---||---}
\texttt{ramage02} & 16830	&	84.66	& \textbf{80909} & \textbf{80909} & 698100 & \textbf{80419} & \textbf{80419} & 689912\\
\hhline{-||--||---||---}
\texttt{pkustk07} & 16860	&	71.23	& \textbf{66699} & \textbf{66699} & 583500 & \textbf{66182} & \textbf{66182} & 514741\\
\hhline{-||--||---||---}
\end{tabular}

\begin{tabular}{|c||c|c|c||c|c|c|}
\multicolumn{7}{c}{~}\\
\hhline{-||---||---}
$k=3$ & \multicolumn{3}{c||}{0\% imbalance} & \multicolumn{3}{c|}{1\% imbalance} \\
\cline{2-7}
graph & DA & KaHIP & Gurobi & DA & KaHIP & Gurobi\\
\hhline{-||---||---}
\texttt{exdata\_1} & \textbf{2668} & 17550 & 2840 & \textbf{2654} & 14576 & 2664\\
\hhline{-||---||---}
\texttt{TSC\_OPF\_1047} & 78608 & \textbf{78369} & 668861 & 72610 & \textbf{72066} & 669320\\
\hhline{-||---||---}
\texttt{nd3k} & \textbf{230329} & \textbf{230329} & 1088853 & 230099 & \textbf{230042} & 1053084\\
\hhline{-||---||---}
\texttt{nemeth26} & \textbf{6748} & \textbf{6748} & 141759 & \textbf{6736} & \textbf{6736} & 122437\\
\hhline{-||---||---}
\end{tabular}
\begin{tabular}{|c|c|c||c|c|c|}
\multicolumn{6}{c}{~}\\
\hhline{---||---}
\multicolumn{3}{|c|}{3\% imbalance} & \multicolumn{3}{c|}{5\% imbalance} \\
\hhline{---||---}
 DA & KaHIP & Gurobi & DA & KaHIP & Gurobi\\
\hhline{---||---}
 \textbf{2628} & 13066 & \textbf{2628} & \textbf{2600} & 16004 & 2678\\
\hhline{---||---}
 72055 & \textbf{71688} & 669320 & 71243 & \textbf{70727} & 669320\\
\hhline{---||---}
 229898 & \textbf{229652} & 1035165 & 229417 & \textbf{229309} & 1077240\\
\hhline{---||---}
 \textbf{6736} & \textbf{6736} & 41057 & \textbf{6736} & \textbf{6736} & 20107\\
\hhline{---||---}
\end{tabular}
\fi
\end{table*}
\begin{table}[htbp]
\centering
\caption{Objective value (number of cut edges) of graphs generated from \texttt{exdata\_1}, $k=2,3$.}
\label{tab:gp_k_2_exdata}
\begin{tabular}{|c||c|c||c|c|c||c|c|c||c|c|c||c|c|c|}
\hhline{-||--||---||---||---||---|}
$k=2$ & \multirow{2}{*}{$|V|$} & \multirow{2}{*}{$d_{avg}$} & \multicolumn{3}{c||}{0\% imbalance} & \multicolumn{3}{c||}{1\% imbalance} & \multicolumn{3}{c||}{3\% imbalance} & \multicolumn{3}{c|}{5\% imbalance}  \\
\hhline{~||~~||---||---||---||---|}
graph &&& DA & KaHIP & Gurobi & DA & KaHIP & Gurobi& DA & KaHIP & Gurobi& DA & KaHIP & Gurobi\\
\hhline{-||--||---||---||---||---|}
\texttt{exdata\_2} & 6016	&	184.22	& \textbf{31769} & 32716 & 554144 & \textbf{31319} & 32373 & 554060& \textbf{30421} & 31536 & 554060 & \textbf{29554} & 30462 & 554060\\
\hhline{-||--||---||---||---||---|}
\texttt{exdata\_3} & 6225	&	187.60	& \textbf{32002} & 32841 & 583986 & \textbf{31573} & 32527 & 584154& \textbf{30736} & 31790 & 583975 & \textbf{29913} & 30746 & 583975\\
\hhline{-||--||---||---||---||---|}
\texttt{exdata\_4} & 6026	&	172.27	& \textbf{25446} & 26142& 519131 & \textbf{25128} & 26015 & 518990& \textbf{24504} & 25338 & 518990 & \textbf{23904} & 24708 & 518990 \\
\hhline{-||--||---||---||---||---|}
\texttt{exdata\_5} & 6052	&	183.74	& \textbf{28422} & 29466 & 556394 & \textbf{28051} & 29104 & 556311& \textbf{27325} & 28333 & 556311 & \textbf{26593} & 27648 & 556311 \\
\hhline{-||--||---||---||---||---|}
\texttt{exdata\_6} & 6056	&	183.22	& \textbf{658} & 668 & 555451 & \textbf{552} & 564 & 5374& \textbf{310} & 320 & 555398 & \textbf{169} & \textbf{169} & 555398 \\
\hhline{-||--||---||---||---||---|}
\texttt{exdata\_7} & 6028	&	186.86	& \textbf{30612} & 31520 & 563061 & \textbf{30192} & 31118 & 562987& \textbf{29355} & 30285 & 562987 & \textbf{28545} & 29511 & 562987 \\
\hhline{-||--||---||---||---||---|}
\texttt{exdata\_8} & 6074	&	185.45	& \textbf{30682} & 31071 & 563758 & \textbf{30242} & 30996 & 563668& \textbf{29425} & 30112 & 563657 & \textbf{28641} & 29353 & 563657 \\
\hhline{-||--||---||---||---||---|}
\texttt{exdata\_9} & 6046	&	184.24	& \textbf{627} & 655 & 27805 & \textbf{492} & 511 & 557185& \textbf{256} & 268 & 1268 & \textbf{117} & \textbf{117} & 7910 \\
\hhline{-||--||---||---||---||---|}
\texttt{exdata\_10} & 6139	&	192.34	& \textbf{29434} & 30298 & 588719 & \textbf{29059} & 29897 & 588986& \textbf{28296} & 29162 & 588729 & \textbf{27561} & 28591 & 588729 \\
\hhline{-||--||---||---||---||---|}
\texttt{exdata\_11} & 6091	&	130.47	& \textbf{1052} & 1364 & 1054 & \textbf{1033} & 1282 & 1736& \textbf{990} & 1215 & \textbf{990} & 954 & 1197 & \textbf{950} \\
\hhline{-||--||---||---||---||---|}
\texttt{exdata\_12} & 6390	&	169.91	&
\textbf{1290} & 1585 & 2087 & \textbf{1270} & 1509 & 541868& \textbf{1250} & 1395 & 541810 & \textbf{1206} & 1330 & 541810
\\
\hhline{-||--||---||---||---||---|}
\texttt{exdata\_13} & 6026	&	108.19	&
\textbf{813} & 1160 & 820 & 843 & 1118 & \textbf{841}& 799 & 983 & \textbf{781} & 733 & 921 & \textbf{713}
\\
\hhline{-||--||---||---||---||---|}
\texttt{exdata\_14} & 5827	&	99.43	&
610 & 766 & \textbf{608} & 616 & 732 & \textbf{598}& 584 & 668 & \textbf{582} & 582 & 646 & \textbf{560}
\\
\hhline{-||--||---||---||---||---|}
\texttt{exdata\_15} & 6380	&	153.01	&
\textbf{1295} & 1496 & \textbf{1295} & 1277 & 1468 & \textbf{1275}& 1235 & 1368 & \textbf{1232} & \textbf{1189} & 1340 & \textbf{1189}
\\
\hhline{-||--||---||---||---||---|}
\texttt{exdata\_16} & 6686	&	176.23	&
\textbf{838} & 5816 & \textbf{838} & 826 & 5806 & \textbf{816}& \textbf{796} & 1710 & 828 & 752 & 1081 & \textbf{728}
\\
\hhline{-||--||---||---||---||---|}
\texttt{exdata\_17} & 5813	&	118.20	&
1044 & 1220 & \textbf{1043} & 1038 & 1238 & \textbf{1024}& 996 & 1094 & \textbf{985} & 956 & 1117 & \textbf{946}
\\
\hhline{-||--||---||---||---||---|}
\texttt{exdata\_18} & 5769	&	136.95	&
\textbf{1058} & 1288 & 1090 & 1050 & 1317 & \textbf{1048}& 1002 & 1164 & \textbf{1000} & 979 & 1135 & \textbf{962}
\\
\hhline{-||--||---||---||---||---|}
\texttt{exdata\_19} & 6062	&	108.49	&
\textbf{950} & 1041 & 329090 & \textbf{932} & 1003 & 2651& \textbf{890} & 951 & 3973 & 853 & 925 & \textbf{851}
\\
\hhline{-||--||---||---||---||---|}
\texttt{exdata\_20} & 5990	&	159.65	& \textbf{1070} & 1474 & 1093 & \textbf{1051} & 1382 & \textbf{1051}& \textbf{1015} & 1620 & 1025 & \textbf{977} & 1688 & 978
\\
\hhline{-||--||---||---||---||---|}
\end{tabular}

\begin{tabular}{|c||c|c||c|c|c||c|c|c||c|c|c||c|c|c|}
\multicolumn{15}{c}{~}\\
\hhline{-||--||---||---||---||---|}
$k=3$ & \multirow{2}{*}{$|V|$} & \multirow{2}{*}{$d_{avg}$} & \multicolumn{3}{c||}{0\% imbalance} & \multicolumn{3}{c||}{1\% imbalance} & \multicolumn{3}{c||}{3\% imbalance} & \multicolumn{3}{c|}{5\% imbalance}  \\
\hhline{~||~~||---||---||---||---|}
graph &&& DA & KaHIP & Gurobi & DA & KaHIP & Gurobi& DA & KaHIP & Gurobi& DA & KaHIP & Gurobi\\
\hhline{-||--||---||---||---||---|}
\texttt{exdata\_2} & 6016	&	184.22	&
\textbf{47529} & 48008 & 738783 & \textbf{47215} & 47519 & 738784& \textbf{46579} & 47179 & 738784 & \textbf{46005} & 46460 & 738784
\\
\hhline{-||--||---||---||---||---|}
\texttt{exdata\_3} & 6225	&	187.60	&
\textbf{46893} & 47433 & 778088 & \textbf{46649} & 47234 & 778091& \textbf{45911} & 46556 & 778091 & \textbf{45289} & 45844 & 778091
\\
\hhline{-||--||---||---||---||---|}
\texttt{exdata\_4} & 6026	&	172.27	&
\textbf{37468} & 38530 & 692385 & \textbf{37258} & 38365 & 692383& \textbf{36692} & 37649 & 692383 & \textbf{36186} & 37186 & 692383
\\
\hhline{-||--||---||---||---||---|}
\texttt{exdata\_5} & 6052	&	183.74	&
\textbf{42031} & 42749 & 741759 & \textbf{41806} & 42451 & 741761& \textbf{41202} & 41916 & 741761 & \textbf{40552} & 41232 & 741761
\\
\hhline{-||--||---||---||---||---|}
\texttt{exdata\_6} & 6056	&	183.22	&
\textbf{14306} & 15219 & 740057 & \textbf{14193} & 14814 & 740056& \textbf{13794} & 13974 & 740056 & \textbf{13109} & 13217 & 740056
\\
\hhline{-||--||---||---||---||---|}
\texttt{exdata\_7} & 6028	&	186.86	&
\textbf{45890} & 46420 & 749893 & \textbf{45573} & 46123 & 749886& \textbf{45140} & 45530 & 749886 & \textbf{44471} & 44947 & 749886
\\
\hhline{-||--||---||---||---||---|}
\texttt{exdata\_8} & 6074	&	185.45	&
\textbf{44025} & 44489 & 751243 & \textbf{43738} & 44243 & 751254& \textbf{43133} & 43662 & 751254 & \textbf{42560} & 42920 & 751254
\\
\hhline{-||--||---||---||---||---|}
\texttt{exdata\_9} & 6046	&	184.24	&
\textbf{14627} & 15582 & 742520 & \textbf{14752} & 15176 & 742522& \textbf{14032} & 14305 & 742522 & \textbf{13278} & 13635 & 742522
\\
\hhline{-||--||---||---||---||---|}
\texttt{exdata\_10} & 6139	&	192.34	&
\textbf{43057} & 43983 & 787039 & \textbf{42858} & 43600 & 787032& \textbf{42182} & 43010 & 787032 & \textbf{41596} & 42404 & 787032
\\
\hhline{-||--||---||---||---||---|}
\texttt{exdata\_11} & 6091	&	130.47	&
\textbf{1866} & 2214 & 529928 & \textbf{1838} & 2046 & 529927& \textbf{1795} & 2078 & 529927 & \textbf{1745} & 1985 & 529927
\\
\hhline{-||--||---||---||---||---|}
\texttt{exdata\_12} & 6390	&	169.91	&
\textbf{2723} & 3232 & 724215 & \textbf{2742} & 2971 & 724224& \textbf{2534} & 2789 & 724224 & \textbf{2485} & 2801 & 724224
\\
\hhline{-||--||---||---||---||---|}
\texttt{exdata\_13} & 6026	&	108.19	&
\textbf{1499} & 1606 & 1993 & \textbf{1505} & 1605 & 3272& 1579 & \textbf{1561} & 2421 & \textbf{1474} & 1504 & 2470
\\
\hhline{-||--||---||---||---||---|}
\texttt{exdata\_14} & 5827	&	99.43	&
\textbf{1191} & 1369 & 2666 & \textbf{1194} & 1314 & 1872& \textbf{1163} & 1286 & 2784 & \textbf{1133} & 1277 & 1287
\\
\hhline{-||--||---||---||---||---|}
\texttt{exdata\_15} & 6380	&	153.01	&
\textbf{2571} & 3113 & 18380 & \textbf{2520} & 3045 & 5906& \textbf{2416} & 2873 & 3727 & \textbf{2317} & 2816 & 4651
\\
\hhline{-||--||---||---||---||---|}
\texttt{exdata\_16} & 6686	&	176.23	&
\textbf{1585} & 3323 & 1882 & \textbf{1814} & 1825 & 2110& \textbf{1644} & 3237 & 1909 & \textbf{1606} & 1683 & 2268
\\
\hhline{-||--||---||---||---||---|}
\texttt{exdata\_17} & 5813	&	118.20	&
\textbf{1970} & 1993 & 3925 & 1981 & \textbf{1958} & 4096& 1982 & \textbf{1943} & 3619 & \textbf{1915} & 1923 & 3342
\\
\hhline{-||--||---||---||---||---|}
\texttt{exdata\_18} & 5769	&	136.95	&
\textbf{1908} & 2198 & 2878 & \textbf{1976} & 2103 & 2887& \textbf{1842} & 2059 & 3622 & \textbf{1833} & 1920 & 2772
\\
\hhline{-||--||---||---||---||---|}
\texttt{exdata\_19} & 6062	&	108.49	&
\textbf{2125} & 2402 & 3485 & \textbf{2095} & 2371 & 9231& \textbf{2011} & 2347 & 3008 & \textbf{1908} & 2176 & 2502
\\
\hhline{-||--||---||---||---||---|}
\texttt{exdata\_20} & 5990	&	159.65	&\textbf{1844} & 2097 & 2598 & \textbf{1822} & 2075 & 2559& \textbf{1810} & 1934 & 2888 & \textbf{1691} & 1885 & 2876
\\
\hhline{-||--||---||---||---||---|}
\end{tabular}
\end{table}

\begin{figure}[htbp]
  \centering
\begin{subfigure}{.33\textwidth}
  \centering
  \includegraphics[width=1\linewidth]{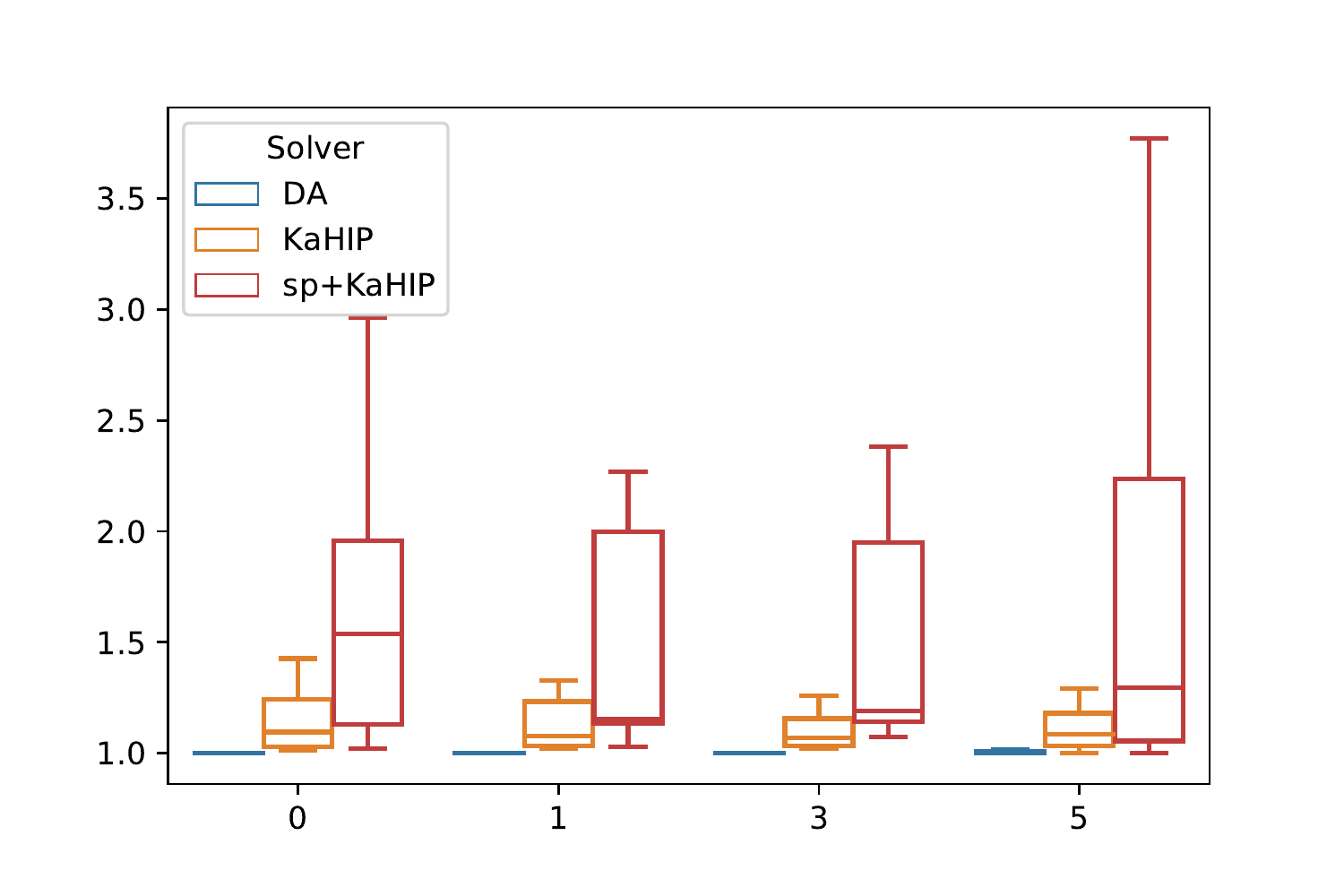}
  \caption{exdata $k=2$}
  \label{fig:sparse1}
\end{subfigure}
\begin{subfigure}{.33\textwidth}
  \centering
  \includegraphics[width=1\linewidth]{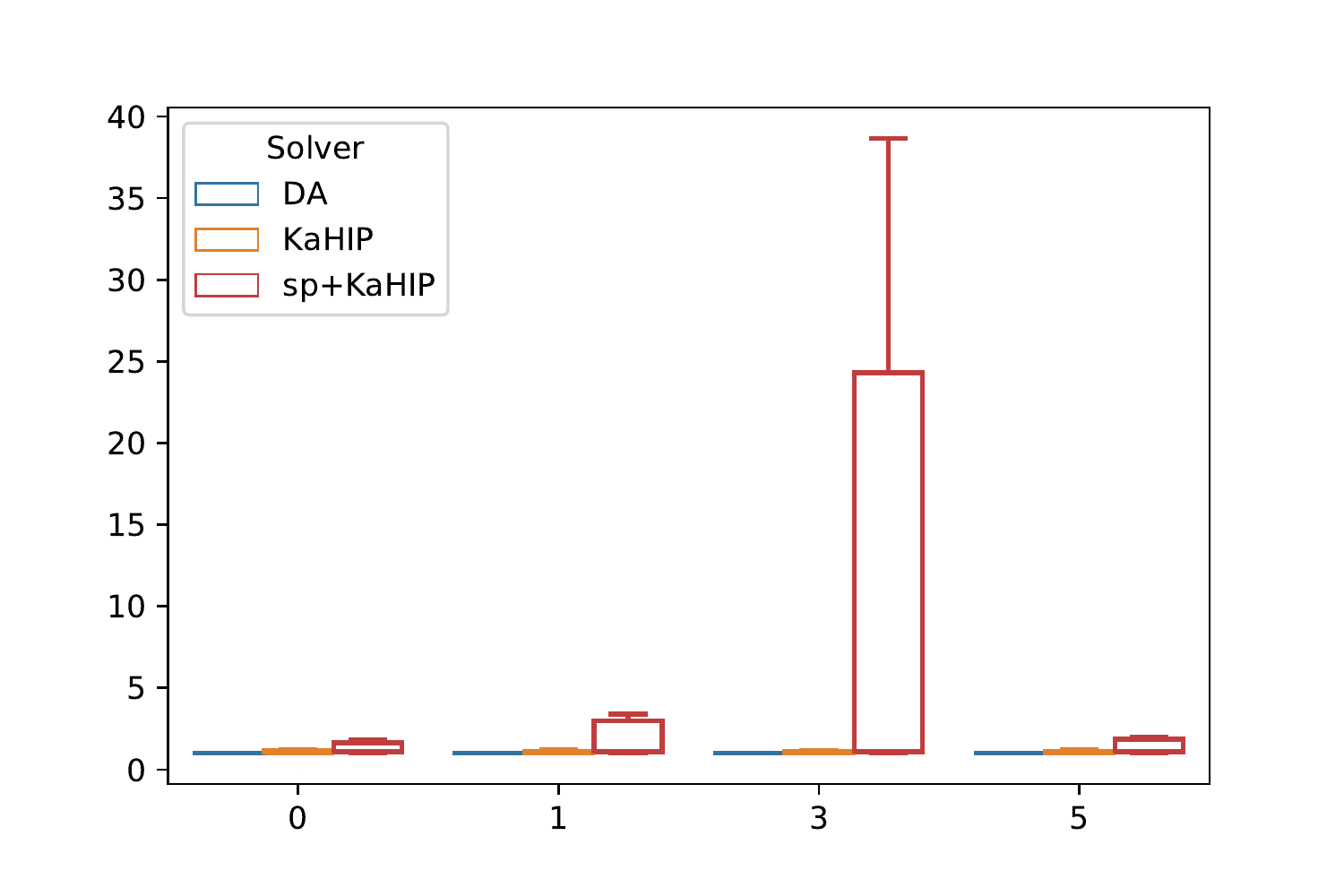}
  \caption{exdata $k=3$}
  \label{fig:sparse2}
\end{subfigure}
  \caption{Comparison of DA, KaHIP and KaHIP with sparsification. The y-axis represents the approximation ratio, the x-axis represents the imbalance factor as percentage}
  \label{fig:sparsify}
\end{figure}

Currently, to tackle GP on dense graphs, the main practical  solution is to first sparsify the graphs
\ifarxiv
\cite{hamann2016structure,safro:spars,spielman2011graph,leskovec2006sampling}
\fi
(hoping that the sparsified graph still preserves the structure of the original dense graph), solve GP on the sparsified graph, and finally project the obtained solution back to the original graph. We have applied  the Forest Fire sparsification  \cite{hamann2016structure} available in Networkit \cite{staudt2016networkit}. This sparsification is based on random walks. The vertices are burned starting from a random vertex, and fire may spread to the neighbors of a burning vertex. The intuition is that the edges that are visited more often during the random walk are more important in the graph. In our experiments, we eliminate  30\% of the edges. Then we solve GP using KaHIP (KaffpaE version) and project the obtained solution back to the original dense graph. We repeat the entire procedure 10 times for each graph, and compare the best results obtained with DA and KaHIP. As shown in Fig. \ref{fig:sparsify}, for dense graphs with complex structures,  KaHIP does not outperform DA, and graph sparsification does not help to achieve this goal. In this case, we advocate the use of the QUBO framework and specialized hardware.

\section{Conclusion and Discussion}
\ifarxiv
As novel technologies for solving computational combinatorial optimization problems emerge, it is important to identify areas in which  these technologies outperform both existing state-of-the-art general-purpose and also problem dedicated solvers. 
\fi
In this work, we have focused on demonstrating practical advantage of software and hardware approaches for the Graph Partitioning problem. We found that dense graphs exhibit limitations of the existing algorithms. 
By experimenting with the Fujitsu Digital Annealer (DA), a quantum-inspired device, we show  graphs on which the DA significantly outperforms current state-of-the-art solvers that are run for identical or longer time. 
\ifarxiv
In particular, we run experiments on instances from three datasets, namely the Walshaw graph partitioning dataset, which represents well-known sparse graphs, graphs from the Sparse-Suite Matrix collection, and lastly synthetically generated graphs. 
\fi
We observe that on sparse graphs (from the Walshaw benchmark) partitioned into two parts with $0\%$ imbalance, the DA returns results identical to the state-of-art graph partitioning solver KaHIP. However, as we increase the imbalance factor and number of parts, we notice that KaHIP outperforms the DA for this dataset. In the Sparse-Suite dataset, we however observe that DA and KaHIP return similar results with a few cases where the DA significantly outperforms KaHIP. Lastly, in our last dataset of synthetically generated graphs, we observe that the DA outperforms KaHIP in almost all cases. With regards to the general-purpose solver, we observe that KaHIP and the DA outperform Gurobi in almost all cases. Our results demonstrate instances where both the DA and KaHIP perform well individually which suggests an opportunity to hybridize  state-of-the-art algorithms and emerging technologies to achieve the best quality/time trade-off. 

\ifarxiv
\section*{Appendix}\label{sec:app}
In this appendix we give the parameters we used with MUSKETEER\footnote{\url{https://github.com/sashagutfraind/musketeer}} \cite{gutfraind2015multiscale} to generate the \texttt{exdata} instances in Table \ref{tab:musketeer}:
\begin{table*}[htb]
\centering
\caption{Parameters used with MUSKETEER}\label{tab:musketeer}
\begin{tabular}{|c||c|c|c|}
\hhline{-||---|}
\texttt{exdata}	&	node growth rate	&	edge edit rate	&	node edit rate	\\
\hhline{-||---|}
2-4	&	[0.01, 0.001]	&	[0.05, 0.04, 0.03]	&	[0.07, 0.06, 0.05]	\\
\hhline{-||---|}
5-7	&	[0.009, 0.001]	&	[0.05, 0.04, 0.03]	&	[0.07, 0.06, 0.05]	\\
\hhline{-||---|}
8-10	&	[0.01, 0.001]	&	[0.05, 0.04, 0.03]	&	[0.07, 0.06, 0.05]	\\
\hhline{-||---|}
11-15	&	[0, 0, 0, 0, 0, 0, 0.01, 0.001]	&	[0, 0, 0, 0, 0, 0.05, 0.04, 0.03]	&	[0, 0, 0, 0, 0, 0.07, 0.06, 0.05]	\\
\hhline{-||---|}
16-20	&	[0, 0, 0, 0, 0, 0, 0, 0.02, 0.002]	&	[0, 0, 0, 0, 0, 0, 0, 0.06, 0.05, 0.04, 0.03]	&	[0, 0, 0, 0, 0, 0, 0, 0.08, 0.07, 0.06, 0.05]	\\
\hhline{-||---|}
\end{tabular}
\end{table*}
\fi
%
%
%

%
%
%
\bibliographystyle{splncs04}
\bibliography{bib}
\end{document}